\documentclass[aps,12pt,tightenlines,longbibliography,notitlepage,floatfix]{revtex4-1}
\pdfoutput=1
\usepackage{amsfonts}
\usepackage{amsmath}
\usepackage{amssymb}
\usepackage{graphicx}
\usepackage{color,bm}
\usepackage{enumitem}
\usepackage[export]{adjustbox}
\usepackage[slantedGreek]{mathptmx}
\usepackage{epstopdf}
\usepackage{latexsym}
\usepackage{epstopdf}

\newcommand{\etal}{\textit{et al.}}

\graphicspath{{images/}}

\begin{document}

\title{Resolvent-based modelling of coherent wavepackets in a turbulent jet}

\author{Lutz Lesshafft$^1$, Onofrio Semeraro$^{1,2}$, Vincent Jaunet$^{3}$, Andr\'e~V.~G.~Cavalieri$^4$ and Peter Jordan$^3$}

\affiliation{
$^1$Laboratoire d'Hydrodynamique, CNRS / \'Ecole polytechnique, 91120 Palaiseau, France
        \\
$^2$LIMSI, CNRS / Universit\'e Paris-Saclay, Orsay, France
		\\
$^3$Institut Pprime, CNRS / Universit\'e de Poitiers / ENSMA, 86962 Futuroscope Chasseneuil, France
        \\
$^4$Divis\~ao de Engenharia Aeron\'autica, Instituto Tecnol\'ogico de Aeron\'autica, S\~ao Jos\'e dos Campos, SP, Brazil}


\begin{abstract}
Coherent turbulent wavepacket structures in a jet at Reynolds number $460\,000$ and Mach number $0.4$ are extracted from experimental measurements, and are modelled as linear fluctuations around the mean flow. The linear model is based on harmonic optimal forcing structures and their associated flow response at individual Strouhal numbers, obtained from analysis of the global linear resolvent operator. These forcing/response wavepackets (`resolvent modes') are first discussed with regard to relevant physical mechanisms that provide energy gain of flow perturbations in the jet. Modal shear instability and the non-modal Orr mechanism are identified as dominant elements, cleanly separated between the optimal and sub-optimal forcing/response pairs. A theoretical development in the framework of spectral covariance dynamics then explicates the link between linear harmonic forcing/response structures and the cross-spectral density (CSD) of stochastic turbulent fluctuations. A low-rank model of the CSD at given Strouhal number is formulated from a truncated set of linear resolvent modes. Corresponding experimental CSD matrices are constructed from extensive two-point velocity measurements. Their eigenmodes (spectral proper orthogonal decomposition or SPOD modes) represent coherent wavepacket structures, and these are compared to their counterparts obtained from the linear model. Close agreement is demonstrated in the range of `preferred mode' Strouhal numbers, around a value of $0.4$, between the leading coherent wavepacket structures as educed from the experiment and from the linear resolvent-based model.
\end{abstract}

\maketitle


\section{Introduction}
\label{sec:intro}

The presence of orderly structures in many turbulent shear flows has been abundantly documented over the last fifty years; in the case of jets, such studies have largely been motivated by the need to reduce their noise generation. It was recognised early on that coherent structures in turbulent jets strongly resemble instability wavepackets, as if they were governed by linear dynamics of small-amplitude fluctuations in a time-averaged mean flow \cite{jordan2013wave}. Many variants of linear analysis techniques have since been explored, in order to identify a model that may faithfully reproduce the coherent turbulence structures in jets. Based on the assumption that linear jet instability is driven by incoming disturbances from upstream, Michalke \cite{michalke1971instability} computed the spatial growth of linear perturbations in parallel jet profiles, followed by the inclusion of weakly non-parallel effects by way of multiple-scales expansion \cite{crighton1976stability,oberleithner2014mean} or parabolised stability equations (PSE, \cite{gudmundsson2011instability}), as well as fully non-parallel linear simulations with inlet forcing \cite{baqui2015coherence}. As discussed by Jordan \& Colonius \cite{jordan2013wave}, all these studies successfully predict the observed spatial growth of coherent turbulence structures near the nozzle, over a dominant but restricted range of frequencies. However, the underlying theoretical model \emph{a priori} pertains to deterministic linear perturbations developing in a steady laminar base flow, and the justification for extending it to chaotic nonlinear fluctuations around a statistical turbulent mean state has remained vague. 

Model equations that govern the statistical moments of turbulent flow, like average and covariance values, can be constructed by considering the linearised Navier--Stokes system subject to stochastic forcing \cite{farrell1993stochastic}; this approach has recently evolved into the `statistical state dynamics' framework, where stochastic forcing of higher-order statistical moments is considered \cite{farrell2014statistical}. Application of this framework to the aerodynamic turbulent jet problem is very promising; the interpretation, given on page 6 in Ref.~\cite{farrell2014statistical}, that ``turbulence in shear flow can be essentially understood as determined by quasi-linear interaction occurring directly between a spatial or temporal mean flow and perturbations'', whereas ``the role of nonlinearity in the dynamics of turbulence is highly restricted'', is clearly born out by the empirical evidence of turbulent jet studies \cite{jordan2013wave}. While most of the literature on stochastic forcing in the linearised Navier--Stokes equations, including the reviews by Schmid (Sec.~4 in \cite{schmid2007}) and Bagheri \etal~\cite{bagheri2009input}, focuses on time-domain formulations of covariance dynamics, coherence in jet turbulence has often been analysed in the frequency domain. In particular, several recent jet studies make use of \emph{spectral proper orthogonal decomposition} (SPOD, see \cite{picard2000pressure}, not to be confused with \cite{sieberSPOD}) as a means to extract empirical coherent structures at a given frequency from experimental or numerical flow data \cite{gudmundsson2011instability,cavalieri2013wavepackets,rodriguez2015}.

Linear instability analysis of jets in recent years has increasingly been carried out in a frequency-domain framework based on \emph{optimal} forcing and associated flow response structures, with no limiting assumptions about the spatial development of the base state \cite{garnaud2013preferred,garnaud2013global,jeun2016input,semeraro2016modeling,semeraro2016stochastic,towne2018,schmidt2018}. Forcing and response structures in this formalism are distributed throughout the interior of the flow, in contrast to the assumption of pure upstream boundary forcing made in most previous models (as cited above), and they are found as the singular modes of the global resolvent operator \cite{schmid2007}. This global resolvent framework, also referred to as `frequency response' \cite{farrell1996generalized,garnaud2013preferred} or `input-output' \cite{jeun2016input} analysis, has similarly been applied in the study of boundary layers \cite{alizard2009sensitivity,monokrousos2010global,sipp2013characterization}, and its potential for the modelling of stochastic dynamics has been explored for backward-facing step flow \cite{dergham2013stochastic,boujo2015sensitivity,beneddine2016conditions}. The question at this point remains, exactly what stochastic quantities can be consistently modelled on the basis of linear resolvent analysis? Dergham \etal~\cite{dergham2013stochastic} use a low-rank resolvent model in order to construct approximations of time-domain POD modes, whereas Boujo \& Gallaire \cite{boujo2015sensitivity} follow the arguments of Farrell \& Ioannou \cite{farrell1996generalized} in order to estimate the frequency spectrum of the stochastic flow response to white-noise forcing. Beneddine \etal~\cite{beneddine2016conditions} go further and set out to model the \emph{spatial distribution} of coherent fluctuations in the frequency domain; they demonstrate convincing agreement between the spatial structures of the optimal linear flow response and the leading SPOD mode, obtained from numerical simulations. 

A formal justification for a direct comparison between optimal linear response structures and SPOD modes has been suggested by Beneddine \etal~\cite{beneddine2016conditions}, and, with an increasing level of detail, in two conference papers \cite{towne2015stochastic,semeraro2016stochastic}, and by Towne \etal~\cite{towne2018}, who also examine the link between resolvent modes and dynamic mode decomposition (DMD). A recent review article \cite{cavalieri2019amr} provides a didactical introduction to resolvent-based modelling of SPOD modes, including numerical codes for a simple model problem, and a discussion of its relevance for the study of jet noise.
Schmidt \etal~\cite{schmidt2018} present a detailed comparison between resolvent analysis results and SPOD modes, extracted from LES data, for high-Reynolds-number turbulent jets at Mach numbers 0.4, 0.9 and 1.5. It is found that the leading SPOD mode is well reproduced by the optimal linear flow response, at the dominant Strouhal number 0.6 for the $Ma=0.4$ case.

The present paper revisits the same turbulent jet configuration, at Mach number 0.4 and Reynolds number $460\, 000$, entirely based on the experimental measurements by Cavalieri \etal~\cite{cavalieri2013wavepackets} and Jaunet \etal~\cite{jaunet2017prf}. The latter study involved velocity measurements in cross-planes of the jet by means of two high-cadence, stereoscopic particle-image velocimetry systems that could be displaced in the streamwise direction so as to provide the cross-spectral density (CSD) of the velocity fluctuations, decomposed both in frequency and in azimuth. In this paper, SPOD modes will be extracted from these experimental CSD matrices, such that they can be compared with linear predictions derived from a resolvent analysis of the experimental mean flow. The principal new aspects of the present study are (i) the use of an experimental database for jet resolvent analysis, (ii) the extraction of SPOD modes from experimental jet measurements, (iii) the design of a \emph{resolvent-based linear model} for such experimental SPOD modes, which are necessarily based on partial-state information, and (iv) a detailed discussion of the linear instability dynamics triggered by optimal and sub-optimal forcing in thin-shear-layer jets, which, by extension, underpin the spectral covariance dynamics contained in the SPOD modes. Although the results of this study are mostly consistent with those of Schmidt \etal~\cite{schmidt2018}, several quantitative as well as qualitative differences arise, with relevance for the physical interpretation in terms of instability mechanisms. These differences are attributed to the inclusion of a nozzle pipe in the present analysis. The nozzle boundary layer is identified as the most receptive flow region in the following calculations, underlining the importance of its numerical resolution, similar to recent observations in large-eddy simulations (LES) \cite{bres2018importance}.

The flow configuration, corresponding to the jet experiments, is briefly defined in Sec.~\ref{sec:setup}. The linear resolvent analysis, including the modal decomposition framework, the numerical implementation, and the presentation of results, is documented in Sec.~\ref{sec:resolvent}. This is followed, in Sec.~\ref{sec:localanal}, by a discussion of the salient linear instability mechanisms that are active in optimal and sub-optimal jet forcing. Section \ref{sec:expvsmodel} presents a detailed comparison between SPOD modes from experimental data and stochastic predictions derived from the resolvent-based linear model. Our new results are then put into perspective with regard to previous modelling attempts. The paper closes, in Sec.~\ref{sec:conclusion}, with a summary of the main conclusions.
%

\section{Flow configuration}\label{sec:setup}

The study is based on jet experiments conducted at the \emph{Bruit et Vent} jet-noise facility of the Pprime Institute in Poitiers. Technical details of the experimental apparatus, as well as measurement validation, are thoroughly described in past publications \cite{cavalieri2013wavepackets,jaunet2017prf}.

The experiments are performed on a $Ma=0.4$ isothermal jet issuing from a convergent-straight nozzle. The Reynolds number of the jet, based on the nozzle exit diameter $D=50$ mm and the maximum exit velocity $U_j$ is defined as $Re=U_jD/\nu=460\,000$, where $\nu$ is the kinematic viscosity. The Strouhal number corresponding to the dimensional frequency $f$ is defined as $St=fD/U_j$. The transition to turbulence of the incoming boundary layer is forced using an azimuthally homogeneous carborandum strip, such that a fully turbulent boundary layer is obtained at the exit section of the nozzle (see Fig.~1 in \cite{cavalieri2013wavepackets}).

\begin{figure}
	\centering
	\includegraphics[width=\columnwidth]{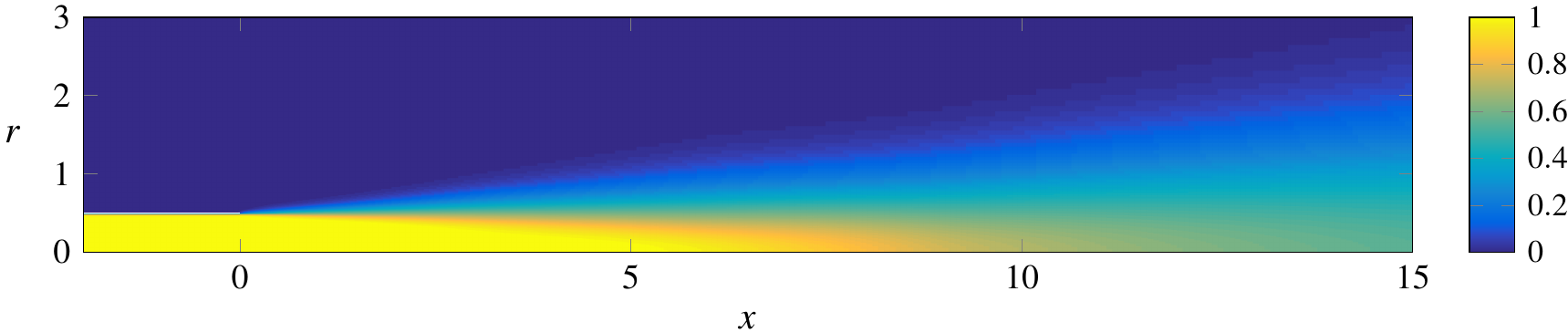}
	\caption{Axial velocity of the mean flow, as used throughout this study \cite{meanflow}. The distribution has been modelled such as to closely reproduce the experimental measurements \cite{cavalieri2013wavepackets}. The pipe wall is represented as a white line, and only a portion of the numerical domain is shown. The rasterisation of the colour plot corresponds to the standard numerical grid resolution (Sec.~\ref{sec:num}).} \label{fig:mean flow}
\end{figure}

\begin{figure}
\centering
\includegraphics[width=\columnwidth]{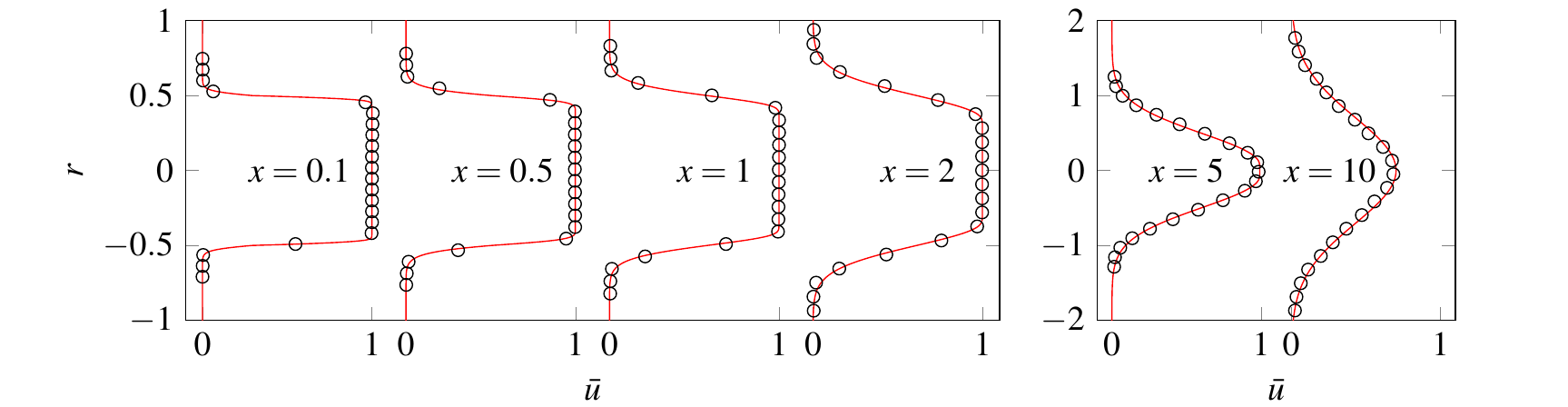}
\caption{ Comparison of axial velocity profiles along $r$, at various $x$ locations, between the numerically modelled mean flow (red lines) and the experimental measurements \cite{cavalieri2013wavepackets} (circles).} \label{fig:meanflowprofiles}
\end{figure}

Free-jet mean flow measurements from this setup, obtained with a Pitot tube, are available from the experiments by Cavalieri \etal~\cite{cavalieri2013wavepackets}, and excellent reproducibility has been demonstrated in the more recent experiments by Jaunet \etal~\cite{jaunet2017prf}. These experimental data are used to construct a parametric model of the mean flow, providing smooth variations of axial and radial velocity, density and temperature \cite{meanflow}. The modelled axial velocity field is shown in Fig.~\ref{fig:mean flow}. In the free-jet region, our modelling procedure follows closely the one described in Rodriguez \etal~\cite{rodriguez2015}, but the present mean flow in our numerical domain has been entirely computed anew from the experimental reference data. It includes a straight pipe that is added upstream of the nozzle exit $x=0$, and the mean flow inside this pipe is taken to be parallel. Twenty-one experimental velocity profiles in the free jet are available between $x=0.1D$ and $x=10D$; some of these are compared in Fig.~\ref{fig:meanflowprofiles} to the modelled mean flow at selected $x$ locations. At $x=6.2D$, the interpolated centreline velocity falls below $0.95\, U_j$, which may be taken to mark the end of the potential core. Downstream of $x=10D$, a self-similar development is assumed, according to the solution discussed in Sec.~$5.2$ of Pope \cite{pope2000turbulent},
\begin{equation}
\bar{u}(r,x) = \frac{\bar{u}(0,x)}{(1+ c_1 \eta^2)^2}\, \quad \text{with~~}\eta = \frac{r}{x-x_0} \text{~~and~~} \bar{u}(0,x) = \frac{c_2}{x-x_0}.
\end{equation}
The parameters $c_1$ and $c_2$, as well as the virtual origin $x_0$, are determined from the available experi\-mental profiles. Parallel nozzle flow, near-nozzle jet, and self-similar far field constitute three distinct flow regions, and smooth transitions between these three are enforced by means of weighted superposition in overlap zones \cite{garnaud2012modes,garnaud2013preferred}. Residual non-smoothness in the derivatives of the reconstructed mean flow is removed by applying high-order explicit filtering. The final axial velocity field (Fig.~\ref{fig:mean flow}) is used for computing temperature and density distributions by way of the Crocco--Busemann relation, and the radial velocity component is recovered from the continuity equation \cite{rodriguez2015}. 

The momentum thickness of the free-jet shear layer, defined at a given $x$ position as 
\begin{equation}
\delta_m = \int_0^{r_{max}} \frac{\bar{\rho}\bar{u}}{\rho_\infty U_j} \left(1-\frac{\bar{u}}{U_j} \right) \text{d}r,
\label{eqn:momThickness}
\end{equation} 
grows linearly in $x$, at a rate of $\text{d}\delta_m/\text{d} x\approx 0.031$. This variation is extrapolated upstream from $x=0.1D$ to the nozzle exit at $x=0$, resulting in a momentum thickness $\delta_m(x=0)=0.0075D$, significantly thinner than in the LES-based study by Schmidt \etal~(\cite{schmidt2018}, their figure 2b). This exit profile is taken between $0\le r \le D/2$ to form the parallel flow inside the nozzle, with a linear decay to zero over the first few discretisation points nearest the wall interior. This parallel flow has a thickness $\delta_m=0.0055D$, when evaluated according to (\ref{eqn:momThickness}) with $r_{max}=D/2$.


\section{Linear resolvent analysis of fluctuations around a mean flow}
\label{sec:resolvent}

\subsection{Governing equations}\label{sec:governing}
We consider the compressible Navier--Stokes equations, in terms of conservative variables $\left(\rho, \rho\mathbf{u}, \rho E\right)$, cast in axisymmetric cylindrical coordinates $\left(x,r\right)$. In the notation of \cite{sandberg2007governing}, these equations are
\begin{subequations}\label{Eq:eq}
	\begin{align}
	\dfrac{\partial \rho}{\partial t} + \nabla\left(\rho \mathbf{u}\right) &= 0, \\
	\dfrac{\partial \rho\mathbf{u}}{\partial t} + \nabla\left(\rho\mathbf{u} \otimes \mathbf{u}\right) &= -\nabla p +\nabla \tau, \\
	\dfrac{\partial \rho E}{\partial t} + \nabla\left(\rho\mathbf{u}E\right) &= -\nabla \mathbf{h} +\nabla \left(\tau \mathbf{u} \right) ,
\end{align}
\end{subequations}
where $\rho$ is density and $\mathbf{u}=(u_x,u_r,0)$ is the velocity vector, with axial and radial components $u_x$ and $u_r$, and with zero azimuthal velocity. In our axisymmetric setting, all quantities are independent of the azimuthal coordinate $\theta$. The total energy $E$ is then defined as
\begin{equation}
E = \dfrac{T}{\gamma(\gamma-1)Ma^2}+\dfrac{1}{2}(|u_x|^2+|u_r|^2),
\end{equation}
with $\gamma=1.4$ the ratio of specific heats. The tensor $\tau$ denotes the molecular stresses, and $\mathbf{h}$ is the heat flux vector. The reference length of the problem is the pipe diameter $D$. The reference velocity is chosen as the centreline velocity $U_j$ at the pipe exit $x=0$, and the reference density is set as the ambient value $\rho_\infty$.  Sutherland's law is used to calculate the viscosity, and the Prandtl number is set to $Pr=0.72$, the standard value for air.

\subsection{Representation as a linear input-output system}\label{sec:linsystem}

The flow variables $q=(\rho,\rho u_x,\rho u_r,\rho E)$ are decomposed into their time-averaged mean and time-dependent fluctuation components, $q(x,r,t)=\bar{q}(x,r) + q'(x,r,t)$.
The governing equations (\ref{Eq:eq}) can then be rewritten in the form 
\begin{equation}\label{eq:system}
\dfrac{\partial q'}{\partial t} -Aq' =f,
\end{equation}
where $A$ is the operator obtained by linearising (\ref{Eq:eq}) around the mean flow, and the vector $f$ contains all remaining nonlinearities in $q'$, i.e.~the fluctuations of the generalised Reynolds stresses \cite{reynolds1972mechanics}, as well as any external forcing at the boundaries of a finite-domain flow problem. The vector $f$ thus contains zero-mean source terms of the continuity, momentum and energy equations.

A Fourier-transform  
\begin{equation}
q'(x,r,t)= \int_{-\infty}^\infty \hat{q}(x,r,\omega)e^{i\omega t} \text{d}\omega, \qquad
f(x,r,t)= \int_{-\infty}^\infty \hat{f}(x,r,\omega)e^{i\omega t} \text{d}\omega,
\end{equation}
leads to the frequency-domain system
\begin{equation}\label{eq:system2}
\hat{q} = \left(i\omega I -A\right)^{-1}\hat{f} = R(\omega)\hat{f},
\end{equation}
where $R$ is the \emph{resolvent} operator \cite{schmid2007}.
As $f$ contains all terms nonlinear in $q'$, the forcing with its Fourier-transform $\hat{f}$ induces an inherent coupling between all frequencies. In order to make use of the system (\ref{eq:system2}) for the purpose of modelling, a closure assumption is required that allows a decoupling of frequencies. Following previous literature \cite{farrell1996generalized,beneddine2016conditions,towne2018,mckeon2010critical}, we choose to simply regard $f$ as an anonymous forcing term, representing any incoming perturbations from the nozzle or the ambient, as well as fluctuations in the nonlinear terms of the momentum and energy equations, but without accounting for its inner structure that makes it dependent on $q'$. Accordingly, we neglect the dependence of $\hat{f}$ at one given frequency on $\hat{q}$ at other frequencies.

One possibility to account for a limited interaction between frequencies lies in the inclusion of turbulent dissipation through small scales in the linear operator $A$, in the form of turbulent viscosity. Indeed, any portion of $\hat{f}$ may be modelled as being linearly dependent on $\hat{q}$, without introducing explicit coupling between frequencies. Some empirical evidence suggests the pertinence of such modelling \cite{tammisola2016coherent,oberleithner2015formation}, and we have used it in the past for the resolvent analysis of turbulent jets \cite{semeraroSandbergStockholm,semeraro2016modeling}, but the procedure requires additional modelling hypotheses and is not pursued here. All computations in this section only account for molecular viscosity at $Re=460\, 000$.

\subsection{Modal decomposition of the resolvent operator}\label{sec:optimaz}

 First attempts to model the global linear response of shear flows to forcing were based on eigenmode decomposition, e.g.~\cite{aakervik2008global}. However, it has generally been realised that \emph{amplifier-type} flow dynamics \cite{huerre1990local} are not adequately described by their spectrum of stable eigenmodes. For jet flows, this case is made by Garnaud \etal~\cite{garnaud2013modal}. Instead, a decomposition approach based on \emph{singular modes} (SVD) is conceptually well-suited.

The following development restates the SVD-based resolvent analysis formalism as it has been applied in numerous past studies, including references \cite{garnaud2013preferred,jeun2016input,semeraro2016modeling,semeraro2016stochastic,towne2018,schmidt2018}. It is presented here in a form that establishes our nomenclature and clarifies the influence of the chosen energy norm.

For a given frequency $\omega$, the resolvent operator provides the mapping between any forcing structure $\hat{f}(x,r,\omega)$ and its linear flow response $\hat{q}(x,r,\omega)$. The common choice for an energy measure in compressible settings is the norm defined by Chu \cite{chu1965},
\begin{equation}\label{eq:norm}
	\|\hat{q}\|^2 = \iint_{\Omega}\left(\bar{\rho}(|u_x|^2+|u_r|^2) +\frac{\bar{p}}{\bar{\rho}}|\hat{\rho}|^2
	+\frac{\bar{\rho}^2}{\gamma^2\left(\gamma-1\right)Ma^4 \bar{p}}|\hat{T}|^2\right) r\, \text{d}r \, \text{d}x,
\end{equation}
which is used in the following computations. Both the forcing and the flow response are measured in this norm, and the spatial integration in both cases is carried out over the entire numerical domain $\Omega$, with the exception of absorbing layers near the outer boundaries (see Sec.~\ref{sec:num}). In discrete form, the norm is expressed by a Hermitian positive-definite matrix $M$, such that $\|\hat{q}\|^2 = \hat{q}^H M \hat{q}$, with a Cholesky factorisation $M=N^HN$. Flow forcing and response are represented by \emph{discrete} complex-valued vectors $\hat{f}$ and $\hat{q}$ in the following
	
The \emph{gain} between input and output energy is then defined as
\begin{equation}\label{eq:raile}
	\sigma^2=
	\frac{\|\hat{q}\|^2}{\|\hat{f}\|^2}=
	\frac{\hat{f}^HR^H M R\hat{f}}{\hat{f}^H M\hat{f}}
	=\frac{\hat{v}^H N^{-1,H} R^H MRN^{-1}\hat{v}}{\hat{v}^H\hat{v}},\quad \text{with~}\hat{v}=N\hat{f},
\end{equation}
which has the form of a Rayleigh quotient, involving the Hermitian operator $N^{-1,H} R^H MRN^{-1}$. Consequently, the eigenvectors $\hat{v}_i$ of this operator are orthogonal, its eigenvalues $\sigma^2_i$ are real positive, and the largest possible energy gain of the linear flow system is given by the largest eigenvalue. The forcing structure that gives rise to an energy gain $\sigma_i^2$ is recovered as $\hat{f}_i=N^{-1} \hat{v}_i$. After normalisation, $\hat{v}_i^H\hat{v}_i=1$, the eigenvectors $\hat{v}_i$ are the columns of the right singular matrix $V$ of the operator
\begin{equation}
\label{eqn:svd}
NRN^{-1} = U\Sigma V^H,
\end{equation}
associated with the singular values $\sigma_i$ as entries in the diagonal matrix $\Sigma$, and with the unique unitary matrix $U$.

The forcing structures $\hat{f}_i$ are the columns of a matrix $F=N^{-1}V$, and the associated flow response structures $\hat{q}_i$ form the matrix $\hat{Q}=RF$. With (\ref{eqn:svd}), it is found that  $\hat{Q}=N^{-1}U\Sigma$, from where it follows that $\hat{Q}^HM\hat{Q}=\Sigma^2$. A normalised response matrix $Q=\hat{Q}\Sigma^{-1}$ is introduced, such that the final identities for our modal resolvent decomposition are recovered:
\begin{subequations}
	\begin{align}
	F^H MF&=Q^H MQ=I,\\
	R &= Q\Sigma F^H M. \label{eqn:resolventdecomp}
	\end{align}
\end{subequations}

The singular values $\sigma_i$ are arranged in descending order, such that the optimal energy gain is given by $\sigma_{max}^2=\sigma_1^2$, arising for the forcing structure $\hat{f}_1$. In the inner-product space defined with the matrix $M$, each vector $\hat{f}_j$ represents the optimal forcing in the subspace that is orthogonal to all leading vectors $\hat{f}_i$ with $i>j$. As a convention, we will refer to a given triple ($\sigma_i,\hat{f}_i,\hat{q}_i$) as the resolvent mode $i$, consisting of the $i^{th}$ gain, forcing mode and response mode. The triple ($\sigma_1,\hat{f}_1,\hat{q}_1$) of `optimal gain', `optimal forcing' and `optimal response' is characterised by the maximum value of $\sigma$, and resolvent modes with $i>1$ are sometimes referred to as `sub-optimals'. Note that `gain' in the following refers to $\sigma$, not to the \emph{energy} gain, given by $\sigma^2$.


\subsection{Matrix-free computation of resolvent modes}
\label{sec:num}

Gain values and associated forcing modes are computed by solving the reformulated eigenvalue problem
\begin{equation}
R^H M R \hat{f}_i = \sigma^2_i M \hat{f}_i,
\label{eqn:evp}
\end{equation}
using the iterative Lanczos method that is provided by the SLEPc library \cite{hernandez2005slepc}. A matrix-free time-stepping method is used in each iteration step, as described in detail by \cite{garnaud2012modes}. Time-stepping needs to be performed both for the solution of a direct system, $a=Rb$, and for the subsequent solution of an adjoint system, $a'=R^Hb'$. The time horizon $t_{max}$ of these calculations must be chosen long enough such that the final periodic flow regime is recovered with sufficient accuracy.
A numerical procedure for the adjoint system is constructed according to the method of Fosas de Pando \etal~\cite{de2012efficient}, which ensures that the complete numerical encoding of the operator in (\ref{eqn:evp}) remains strictly Hermitian; this is an important requirement for the efficiency of the Lanczos algorithm. 

%
%
\begin{figure}
		\includegraphics[width=\textwidth]{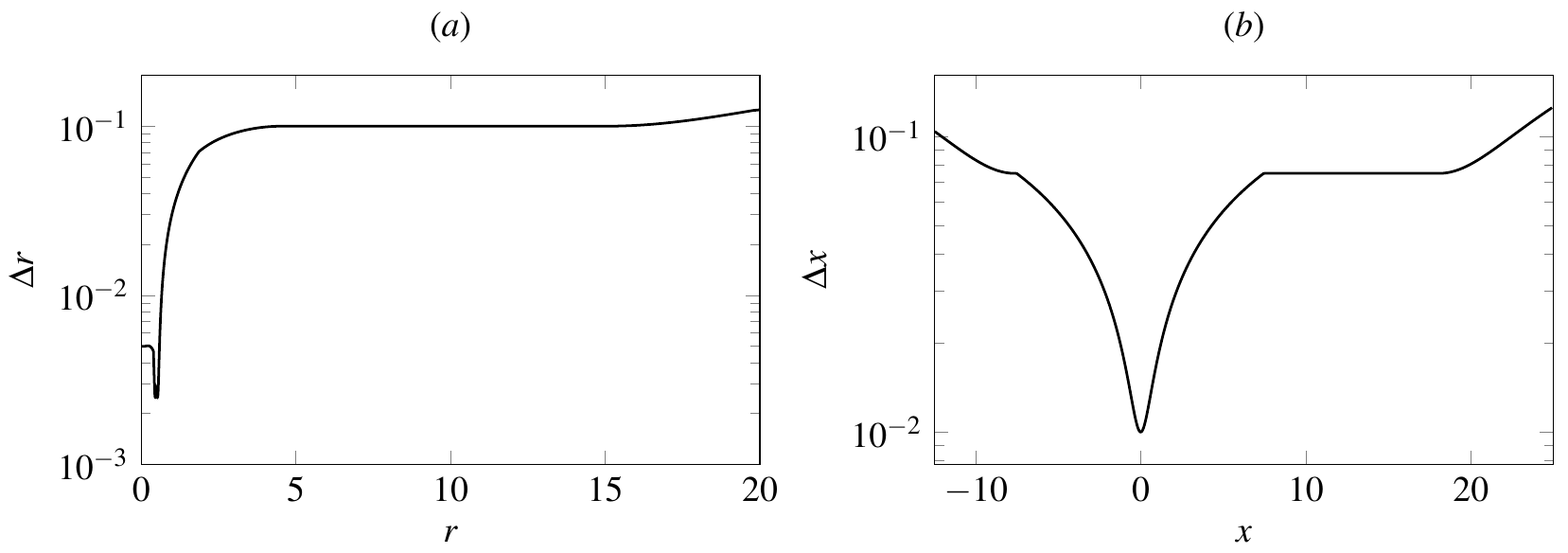}
	\caption{Spacing of mesh points, (a) in the radial direction and (b) in the axial direction. The mesh is orthogonal.}
	\label{fig:mesh}
\end{figure}
%
%
%
\begin{table}
	\centering
	\begin{tabular}{l|cccc}
		\hline
		& $\Delta x_{min}$ & $\Delta r_{min}$ & $t_{max}$ & $\sigma_1$\\
		\hline 
		
		case A    & 0.0200     & 0.0075      &  60       &     5117.2 \\
		case B    & 0.0200     & 0.0025      &  60       &     5256.3\\
		case C0   & 0.0100     & 0.0025      &  60       &     5264.3\\
		case C1   & 0.0100     & 0.0025      &  70       &     5294.4\\
		case C2   & 0.0100     & 0.0025      &  75       &     5302.6\\
		case C3   & 0.0100     & 0.0025      &  80       &     5306.6\\
		\hline 
	\end{tabular}
	\caption{Convergence of the optimal gain $\sigma_1$ at $St=0.5$, as a function of grid resolution and final simulation time $t_{max}$. These test calculations were performed with a reduced Krylov space dimension $N_{kr}=4$. A value $N_{kr}=12$ is used in all following computations for an increased accuracy of sub-optimal modes.}
	\label{tab1}
\end{table}

The linear system (\ref{eq:system}) is discretised with explicit finite-difference schemes \cite{berland2007high}, using an $11$-point stencil.  Time integration is performed with a third-order Runge--Kutta algorithm, with time step $\Delta t= 2.85 \cdot 10^{-3}$. The computational domain extends along the streamwise direction over the interval $x\in [-12.5,\, 25]$, and from the symmetry axis $r=0$ outwards to $r=20$; the nozzle exit is placed at $x=0$. The mesh that is used in all calculations presented in the following sections consists of $(N_x, N_r) = (750,380)$ discretisation points. These points are distributed on a non-uniform Cartesian grid, with maximum resolution along the pipe walls, in the shear layer and around the nozzle lip. Figure~\ref{fig:mesh} displays the axial and radial point distributions.

Symmetry boundary conditions are imposed on the jet axis by the use of ghost points: $\rho$, $\rho u_x$ and $\rho E$ are prescribed to be even functions in $r$ across the axis, while $\rho u_r$ is odd. On all other boundaries, the LODI boundary conditions are applied \cite{poinsot1992}, in combination with absorbing layers \cite{coloniusARFM} at $r>16$, $x<-8$ and $x>21$.

Convergence of the optimal gain is tested with respect to the grid spacing and to the final time $t_{max}$ of the simulations. Several results are reported in table \ref{tab1}. For a fixed value $t_{max}=60$, the mesh of case C0 is deemed sufficiently refined; this is the standard mesh displayed in Fig.~\ref{fig:mesh}. The final time is chosen by tracking the energy of time-harmonic fluctuations, in order to evaluate to what extent transient dynamics have died out. Satisfactory convergence is reached at $t_{max}=80$, which corresponds approximately to twice the convection time of vortical structures between the nozzle exit and the downstream end of the physical domain; this value is retained for all following calculations.

For any given Strouhal number, the five leading resolvent modes are computed, using a Krylov space of dimension $N_{kr}=12$. The Lanczos iteration is halted when the estimated residual norms of all five modes have fallen below the tolerance value $\epsilon=10^{-4}$ (see \cite{hernandez2005slepc}), implying confidence in the first four significant digits of the gain values. A typical computation for one Strouhal number requires about $12$--$16$ wall-time hours on 192 cores of Intel Xeon E5-2690 v3 CPUs.


\subsection{Resolvent mode results}\label{sec:numresults}

\begin{figure}
	\centering
	\includegraphics[width=0.6\textwidth]{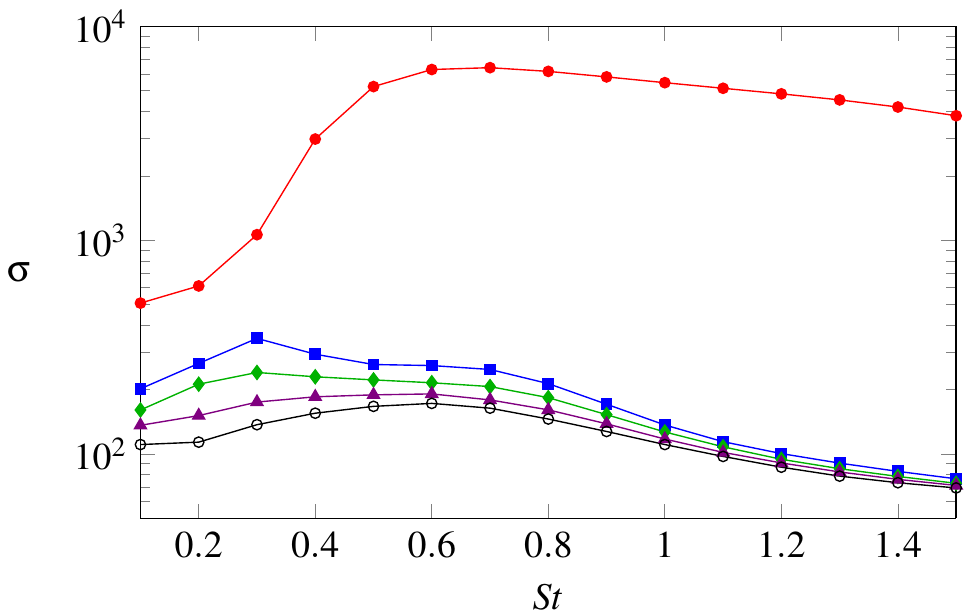}
	\caption{The five leading resolvent gain values ({\color{red}$\bullet$} $\sigma_1$, {\scriptsize\color{blue}$\blacksquare$} $\sigma_2$, {\small\color[rgb]{0,0.7,0}$\blacklozenge$} $\sigma_3$, {\color[rgb]{0.5,0,0.5}$\blacktriangle$} $\sigma_4$, $\circ$ $\sigma_5$) as functions of Strouhal number.}
	\label{fig:gains}
\end{figure}

Gain values of the five leading resolvent modes are shown in Fig.~\ref{fig:gains} as functions of the Strouhal number. 
Above $St=0.3$, the optimal gain curve is well separated from the sub-optimal ones. The maximum overall gain occurs at $St=0.7$, where $\sigma_1$ is one order of magnitude larger than $\sigma_2$. 

%
%

\begin{figure}[p]
	\centering
	\includegraphics[width=\textwidth]{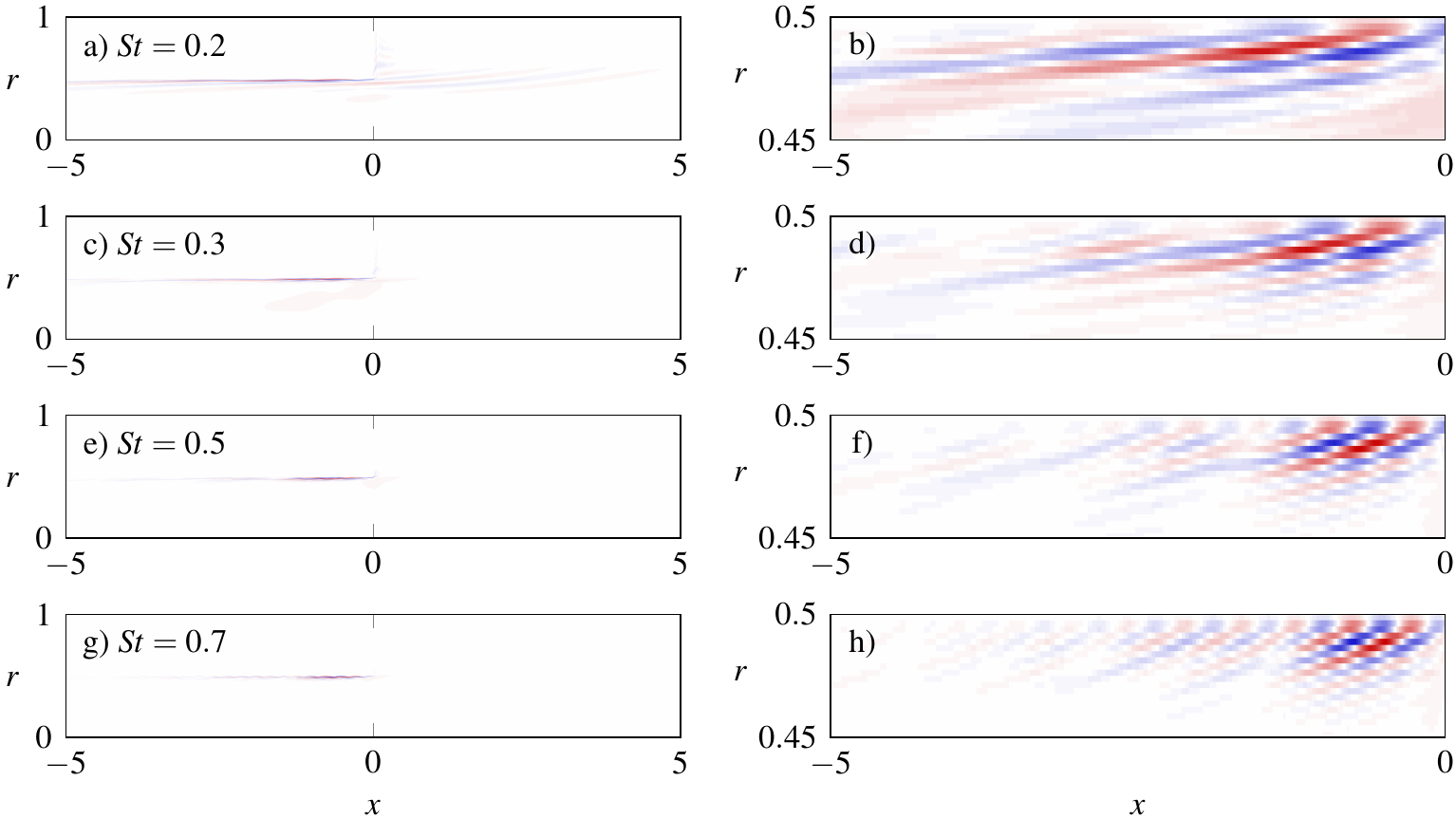}
	\includegraphics[width=0.2\textwidth]{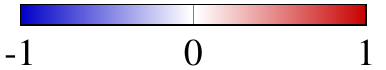}
	\caption{Optimal forcing modes at various Strouhal numbers, associated with the response modes in Fig.~\ref{fig:real_r_opt}. The real part of axial velocity forcing is represented. (a,c,e,g) Optimal forcing, plotted with aspect ratio 2. $(b,d,f,h)$ Close-up of the pipe boundary layer, where the forcing is localised, at the same $St$ values as in the left column. The rasterisation corresponds to the numerical mesh; each field is normalised with respect to its maximum amplitude. \label{fig:real_f_opt}
	}
\end{figure}

\begin{figure}[p]
	\centering
	\includegraphics[width=\textwidth]{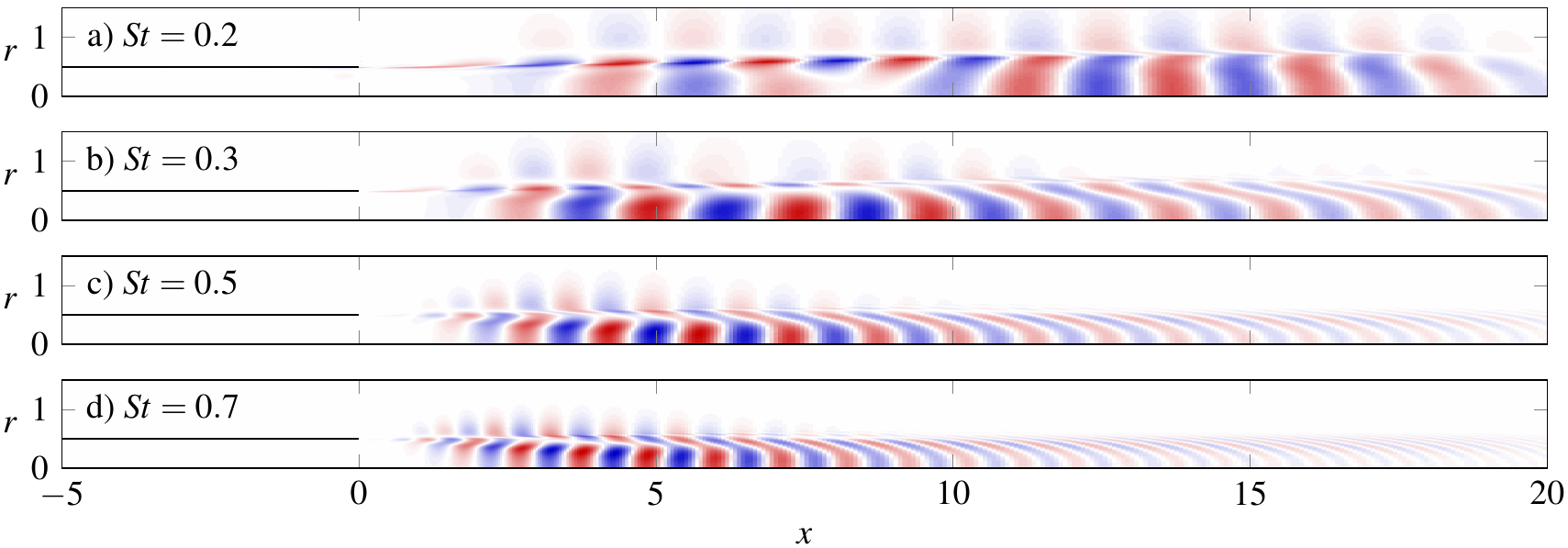}
	\\[4pt]
	\includegraphics[width=0.2\textwidth]{colorbar}
	\caption{Optimal response modes at various Strouhal numbers, associated with the forcing modes in Fig.~\ref{fig:real_f_opt}. The real part of axial velocity perturbations is represented, with aspect ratio 1. The pipe wall is shown as a black line. Each field is normalised with respect to its maximum amplitude. \label{fig:real_r_opt}
	}
\end{figure}

The optimal forcing and response structures are presented in Figs.~\ref{fig:real_f_opt} and \ref{fig:real_r_opt} for several Strouhal numbers between 0.2 and 0.7; snapshots of the real axial momentum components are shown in all frames, with a rasterisation that corresponds to the numerical mesh. At Strouhal numbers 0.3 and above, the forcing is localised in a thin layer at the inner pipe wall, near the nozzle. The right-column frames in Fig.~\ref{fig:real_f_opt} give a magnified view of the forcing in this flow region. Elongated structures are tilted against the flow direction, in a fashion that is typical of the Orr mechanism (see Sec.~\ref{sec:localanal} for a brief description of this phenomenon). Similar optimal forcing structures have been identified in boundary layers \cite{alizard2009sensitivity,monokrousos2010global} and in past studies of incompressible as well as compressible jets \cite{garnaud2013preferred, garnaud2013global,semeraro2016modeling}. The response structures at $St\ge 0.3$ exhibit the classical wavepacket shape associated with shear instability, with peak amplitudes inside the potential core \cite{garnaud2013preferred,rodriguez2015,schmidt2018}. 

The main characteristics of both the optimal forcing and the optimal response modes are similar at all Strouhal numbers above 0.2: optimal forcing acts upstream in the pipe and generates a wavepacket with amplitude growth in the potential core region of the jet. As the Strouhal number increases, the wavelength shortens, and the location of the peak amplitude moves closer to the nozzle, consistent with the interpretation of local spatial instability \cite{michalke1971instability}. At low Strouhal numbers, as shown in figures \ref{fig:real_f_opt}$a$ and \ref{fig:real_r_opt}$a$, different effects seem to arise: in addition to the described scenario, tilted forcing structures protrude into the free shear layer close to the nozzle, and the response wavepacket appears to be composed of two distinct regions. Along the jet axis, one local amplitude maximum occurs at $x=5.5$, and another one at $x=13$, far downstream of the potential core. The low-$St$ optimal mode results of Schmidt \etal~(Fig.~12$f$ in \cite{schmidt2018}) show a similar pattern. As argued by those authors, the distinct mode characteristics at low Strouhal numbers are likely to be associated with a crossing or merging of mode branches, due to a lessened efficiency of the shear instability mechanism.

%
%
%
%
\begin{figure}
	\centering
	\includegraphics[width=\textwidth]{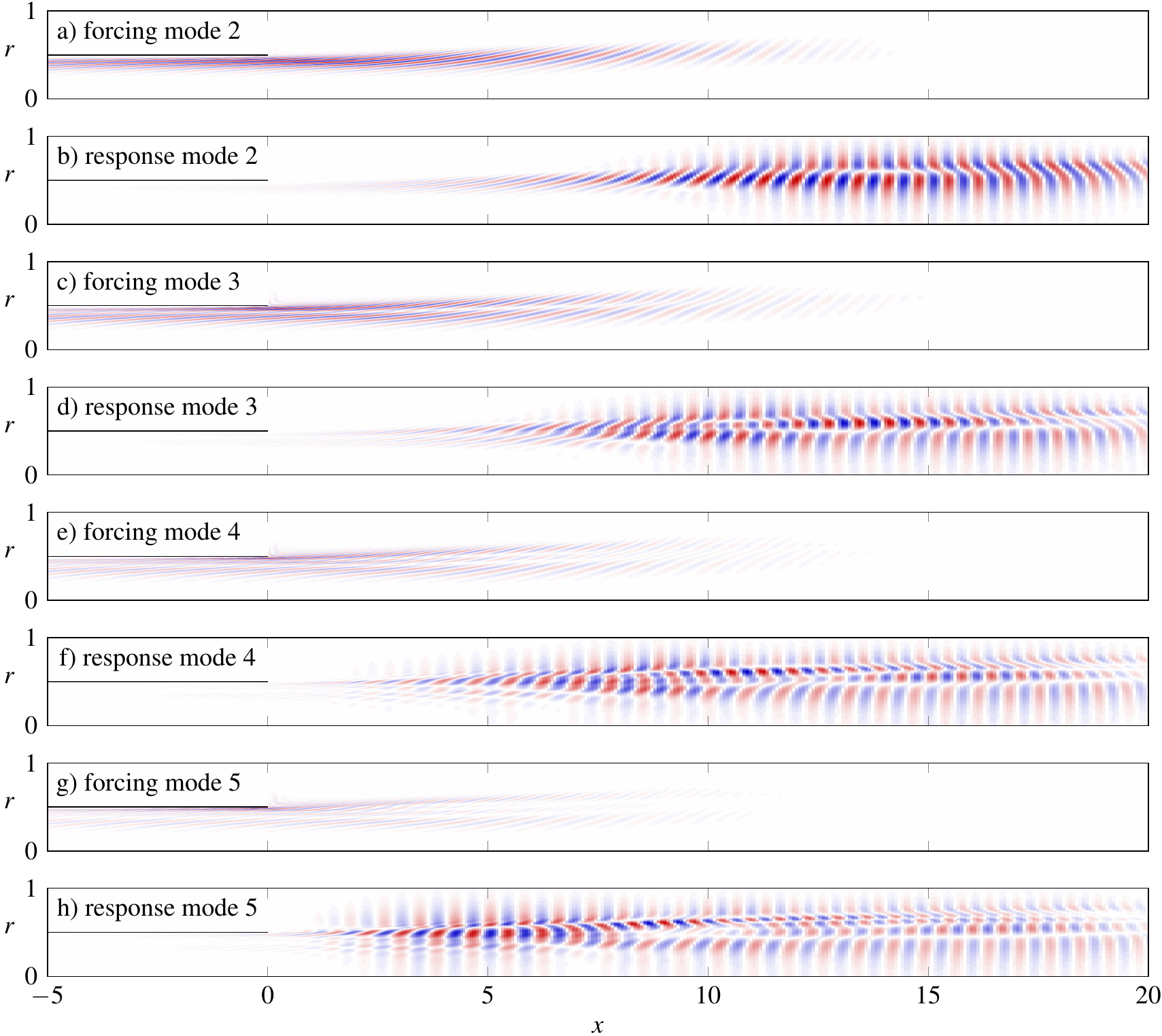}
	\\[4pt]
	\includegraphics[width=0.2\textwidth]{colorbar}
	\caption{Sub-optimal forcing and response modes for ${St=0.7}$. The real part of axial velocity perturbations is represented. (a,c,e,g) Forcing modes; and (b,d,f,h) associated response modes. The pipe wall is shown as a black line. Each field is normalised with respect to its maximum amplitude. The aspect ratio is 2, strong magnification is required in order to visualise the fine-scale radial variations. \label{fig:real_fr_St07}
	}
\end{figure}
Sub-optimal forcing and response structures, modes 2-5, are displayed in Fig.~\ref{fig:real_fr_St07} for $St=0.7$. The forcing in all cases is again characterised by structures that are tilted against the mean flow, although these structures arise at a small radial distance away from the pipe wall, and they extend far into the free jet, with significant amplitude inside the shear region. The associated response wavepackets have their maximum amplitude far downstream of the potential core. The forcing structures display radial variations that are suggestive of orthogonal functions, with an increasing number of zero-amplitude nodes along $r$, and corresponding radial structures are imparted to the response wavepackets. A similar hierarchy of optimal modes has been described in Ref.~\cite{semeraro2016modeling}.

Preliminary results, pertaining to the same flow configuration of the jet experiments, have been presented in a previous conference paper \cite{semeraro2016stochastic}. While all forcing modes in that paper are visually identical to the present results, the associated sub-optimal response modes were quite different, all bearing a strong resemblance to the optimal mode. Those earlier calculations were clearly affected by spurious numerical noise, which in all cases triggered the optimal mode sufficiently so as to overwhelm the true sub-optimal response. Non-smoothness in the base flow, as used in Ref.~\cite{semeraro2016stochastic}, was identified to cause this spurious effect. It has been carefully verified that forcing and response modes in the present results form orthogonal sets, with respect to our scalar product, within the accuracy imposed by the residual tolerance of the Lanczos algorithm.


\section{Interpretation of optimal growth mechanisms}\label{sec:localanal}

The role of modal shear (Kelvin--Helmholtz) and non-modal Orr mechanisms for optimal and sub-optimal jet resolvent modes has often been invoked in the literature (see for instance \cite{tissot2016sensitivity,tissot2017wave,schmidt2018}). The aim of this section is to substantiate this interpretation by use of local analysis (for shear) and a parallel model flow (for Orr).

The `Orr mechanism' denotes a linear phenomenon of vorticity convection in a sheared mean flow that gives rise to a growth of perturbation energy. Vortical perturbation structures of alternating sign, initially tilted at an angle opposite to the mean shear, are convected by the mean flow in a way that their tilting angle is first reduced, until the vortex structures are aligned perpendicular to the main flow direction. This deformation is accompanied by an algebraic energy growth of the perturbations. Subsequently, the structures are tilted further, such that they are increasingly aligned with the main flow direction; this phase is accompanied by energy decay. An example of the Orr-scenario in a parallel jet is discussed in the second half of this section. Butler \& Farrell \cite{butler1992three} investigate the mechanisms of energy transfer between perturbations and the mean flow for this phenomenon, based on the Reynolds--Orr equation, in the context of an initial perturbation that evolves in time. They interpret the energy growth as being caused by an interaction of mean shear and perturbation Reynolds stresses, whereas Jim\'enez \cite{jimenez2013} describes it as being the result of mass conservation.

\begin{figure}
	\centering
	\includegraphics[width=0.6\textwidth]{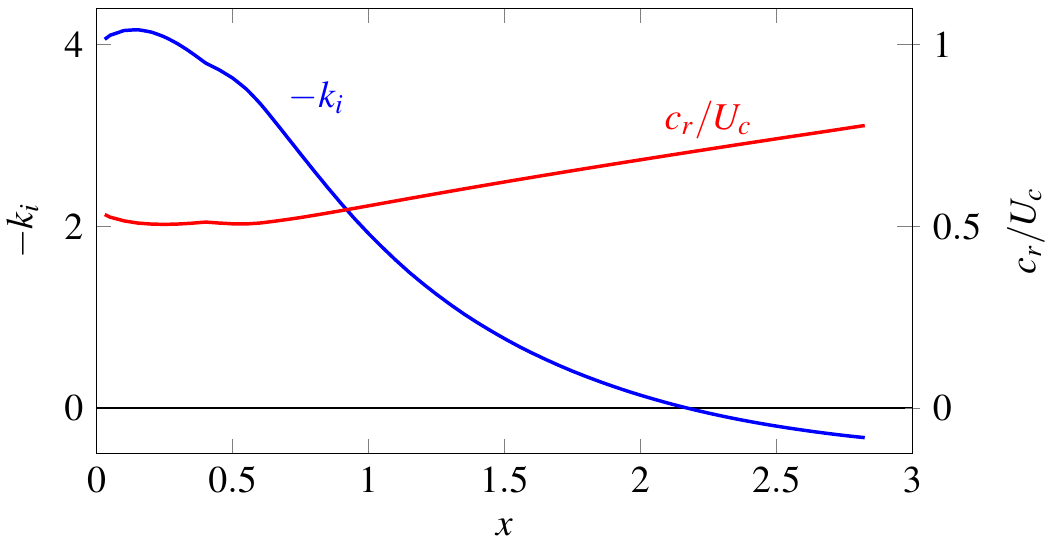}
	\caption{Growth rate $-k_i$ and real phase velocity $c_r$ of the local shear instability mode, and their downstream variations in the jet mean flow. The phase velocity is scaled with the local mean centreline velocity.}\label{fig:KHmode}
\end{figure}

Our optimal resolvent modes (Fig.~\ref{fig:real_r_opt}) strongly resemble those described by Garnaud \etal~\cite{garnaud2013preferred} for an incompressible turbulent jet, which have been interpreted as a constructive combination of the Orr mechanism in the pipe boundary layer and the shear mechanism in the free jet. Close to the nozzle, the optimal response modes display peak amplitudes inside the free shear layer, when measured along the radial direction. As discussed in Ref.~\cite{garnaud2013preferred}, and consistent with many other studies on jet wavepackets (e.g.~\cite{rodriguez2015,schmidt2018}), the spatial distribution as well as the strong streamwise amplitude growth indicates a preponderant role of shear instability in the optimal forcing response. This hypothesis is easily validated by a comparison with local instability results in the near-nozzle region. In a local framework, the shear instability mechanism gives rise to a single spatial $k^+$ mode, which is indeed the only unstable spatial mode that can be found in the jet \cite{lesshafft2007linear}. The downstream evolution of this $k^+$ eigenvalue, for $St=0.7$ in the present jet mean flow, is displayed in Fig.~\ref{fig:KHmode} in terms of its spatial growth rate $-k_i$ and its real phase velocity $c_r=-\omega /k$. The latter is further scaled with the \emph{local} centreline velocity $U_c(x)$ of the jet profile.
%
%

The local shear instability mode, for $St=0.7$, is seen to be unstable only over the interval $0\le x \le 2.17$. Downstream of this position, its eigenfunction (not shown) develops strong oscillations around $r=0.5$, characteristic of the viscous solution in the Stokes sector above the critical point \cite{le1995viscous}, and it remains numerically tractable with confidence over only a short distance further. A Reynolds number of $20\,000$ has been used in these local calculations, lower than in the reference experiment and in the global resolvent analysis, in order to accommodate an accurate resolution of eigenfunctions in the slightly stable regime. It can be demonstrated that results in the \emph{unstable} regime are unaffected by this large value of $Re$. Relevant details on local spectra of compressible jets are discussed by Rodriguez \etal~\cite{rodriguez2015}; in particular, it is described how the shear layer mode, once it is stable, quickly merges into a continuous branch of oscillating modes. The same observations apply here.  
%
%
\begin{figure}
	\centering
	\includegraphics[width=\textwidth]{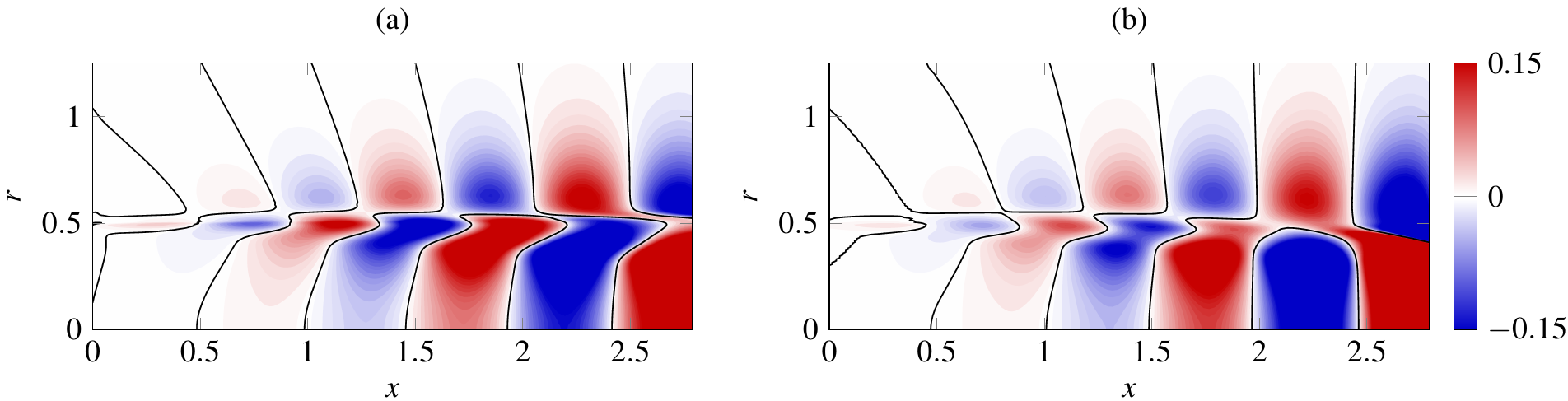}
	\caption{Local shear instability contribution to the optimal response mode at $St=0.7$. (a) Optimal response mode from global resolvent analysis; (b) its reconstruction from projection onto the $k^+$ local shear instability mode. Axial velocity fluctuations are shown over the interval in $x$ where the local mode can be identified numerically. Both fields are normalised with respect to their amplitude maxima, but the colour scale is saturated, in order to make small-amplitude fluctuations visible. The zero contour is traced in black.}\label{fig:KHprojection}
\end{figure}
Following the method of Rodriguez \etal~\cite{rodriguez2015}, the optimal resolvent response wavepacket at $St=0.7$ (Fig.~\ref{fig:KHprojection}$a$) is projected at each $x$ onto the complete basis of local spatial eigenfunctions, by an inner product with the associated local adjoint modes. Projection coefficients are thus obtained at each $x$ position, and the superposition of all local modes indeed fully reproduces the entire global response wavepacket. The isolated contribution of the local $k^+$ shear mode is shown in Fig.~\ref{fig:KHprojection}$b$ over the streamwise region where this mode is identifiable without ambiguity.  It is seen that this mode alone accounts rather accurately for the optimal resolvent response in the unstable interval $0\le x \le 2.17$. Downstream of this region, the radial distribution of the local eigenfunction differs notably from the global result and, contrary to the discussion in Ref.~\cite{garnaud2013preferred}, no other single local mode can be identified as being dominant anywhere for $x>2.17$. The global structure in that region projects significantly onto a large number of local modes from a continuous branch, with strong non-orthogonal cancellation effects.

The second resolvent response structure (mode 2) at $St=0.7$ (Fig.~\ref{fig:real_fr_St07}\textit{b}) cannot be related to any dominant local mode anywhere along $x$. As perturbation growth in the resolvent mode is observed down to a streamwise station $x=14$, it is already obvious that this behaviour is not attributable to modal growth in a local sense, since local instability at this Strouhal number is confined to $x<2.17$. Instead, the spatial features of the response wavepacket suggest again an action of the Orr mechanism, both inside the pipe and in the jet, which feeds on energy gain from the pure convection of tilted vortical structures in a sheared base flow. Such tilted structures are generated by distributed forcing in the free shear layer (Fig.~\ref{fig:real_fr_St07}\textit{a}) upstream of the response maximum. This mechanism has been described by Tissot \etal~\cite{tissot2016sensitivity} in a different framework, where PSE and its adjoint are used to determine forcing terms that optimally match experimental results for the same jet as analysed here. A discussion in terms of local instability modes is not helpful in this case, but the sub-optimal forcing mechanism can still be characterised in the setting of a \emph{parallel} jet flow, which will serve as a model problem in order to understand the trends obtained in the non-parallel framework.

A parallel incompressible jet is considered, defined by a Gaussian velocity profile 
\begin{equation}
U(r) = e^{-2r^2},
\end{equation}
as a simple analytical model for the flow downstream of the potential core. The inflection point is located at $r=0.5$, and the flow is locally stable at a Reynolds number $Re=20\, 000$. Axisymmetric linear perturbations are computed in a numerical domain of $10$ diameters in the axial and radial directions, in response to forcing of both velocity components, which may act anywhere in the domain. The numerical method of \cite{lesshafft2018artificial} is adapted for the \emph{global} computation of optimal resolvent structures, such that the kinetic energy of the flow response at $x=10$, integrated in $r$, is maximised. Thus, forcing is allowed to act throughout the flow domain, but the optimisation objective is measured only at the downstream end.

%
%
\begin{figure}
	\centering
	\includegraphics[width=\textwidth]{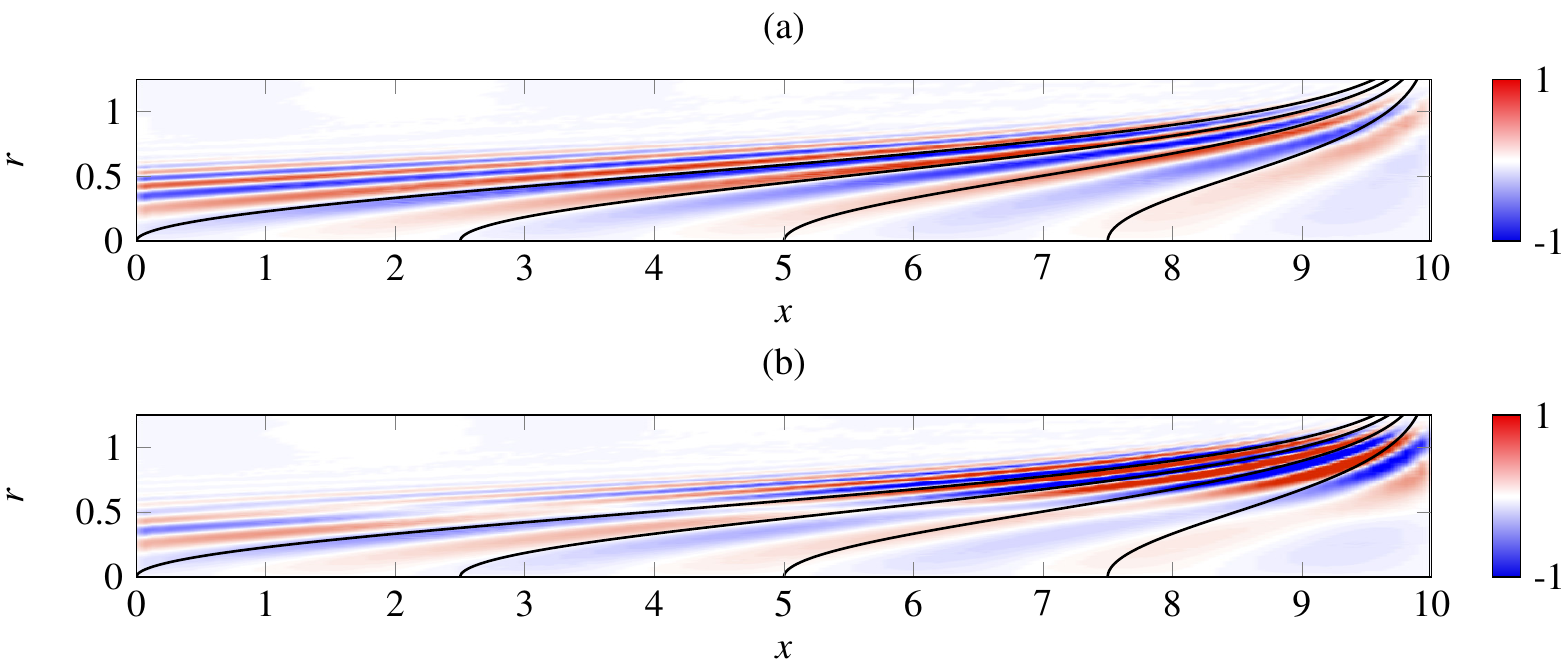}
	\caption{Forcing mode structures in a parallel incompressible jet with Gaussian base flow profile. Kinetic energy at $x=10$ is maximised. Streamwise velocity forcing is shown in linear colour scale. Black lines: contours that are convected into vertical lines at $x=10$ after $\Delta t=(2.5, 5, 7.5, 10)$. a) Forcing mode 1; b) forcing mode 2.}
	\label{fig:parallel_optimal}
\end{figure}
The first two forcing modes are shown in Fig.~\ref{fig:parallel_optimal}, for a Strouhal number $St=1/\pi$. Tilted structures are observed, quite similar to the compressible results displayed in Fig.~\ref{fig:real_fr_St07}(\textit{a,c}), and the tilting angle increases with upstream distance from the target position $x=10$. Black curves trace material lines in the flow that are transported, through convection with the local flow velocity, into vertical lines at $x=10$ after various time lapses. It is immediately seen that the forcing structures follow closely the local curvature of these contours, especially at far upstream positions. Along the black curves, the phase of the first forcing mode structure is approximately constant, whereas the second mode displays a sign change in the phase at the inflection point $r=0.5$. This radial sign change provides for the orthogonality between different forcing modes, and their associated flow responses, while the streamwise variations in modes 1 and 2 are nearly identical. The third and fourth forcing modes of the parallel incompressible jet, not shown in Fig.~\ref{fig:parallel_optimal}, are merely characterised by additional phase changes in the radial direction. All these features are fully consistent with the sub-optimal forcing structures found for the non-parallel compressible jet (Fig.~\ref{fig:real_fr_St07}). The parallel jet results clearly demonstrate that an Orr-type convection mechanism is responsible for the forcing gain in this locally stable setting. From their resemblance, it is inferred that the same mechanism accounts for the gain of sub-optimal structures in the non-parallel compressible jet.

In summary, the following main interpretations of the findings in Sec.~\ref{sec:numresults} are proposed: (i) The optimal forcing (mode 1) targets the shear instability of the jet, leading to exponential amplitude growth along $x$ in a finite region directly downstream of the nozzle. Forcing of this mechanism is most efficient at the upstream end of the locally unstable region. Even more efficient than direct forcing of shear-instability perturbations at the nozzle lip is the forcing of Orr structures in the inner pipe boundary layer, which experience growth before they enter the free jet (consistent with \cite{garnaud2013preferred,garnaud2013global,semeraro2016modeling}). (ii) Sub-optimal forcing exploits the Orr mechanism in the free jet as a means of perturbation energy growth, independent of modal shear instability. Successive sub-optimal modes exhibit an increasing number of sign changes in the phase along $r$, which accounts for their mutual orthogonality. (iii) In our present results (Sec.~\ref{sec:numresults}), the above two mechanisms appear to be well separated in the optimal and sub-optimal resolvent modes at moderate Strouhal numbers. At $St=0.2$, the shape of the optimal mode (figures \ref{fig:real_f_opt}$a$ and \ref{fig:real_r_opt}$a$) suggests a mixed excitation of shear instability and free-jet Orr mechanism.


\section{Coherent structures in jet turbulence: experiment and linear model}
\label{sec:expvsmodel}

The resolvent analysis of Sec.~\ref{sec:resolvent} so far only describes the linear flow response to harmonic forcing input. In this section, those results will be leveraged for the modelling of coherent turbulent structures, where both the forcing $f$ and the response $q'$ are of a stochastic nature.

\subsection{Extraction of SPOD modes from experimental data}
\begin{figure}
	\centering
	\includegraphics[width=0.400\columnwidth]{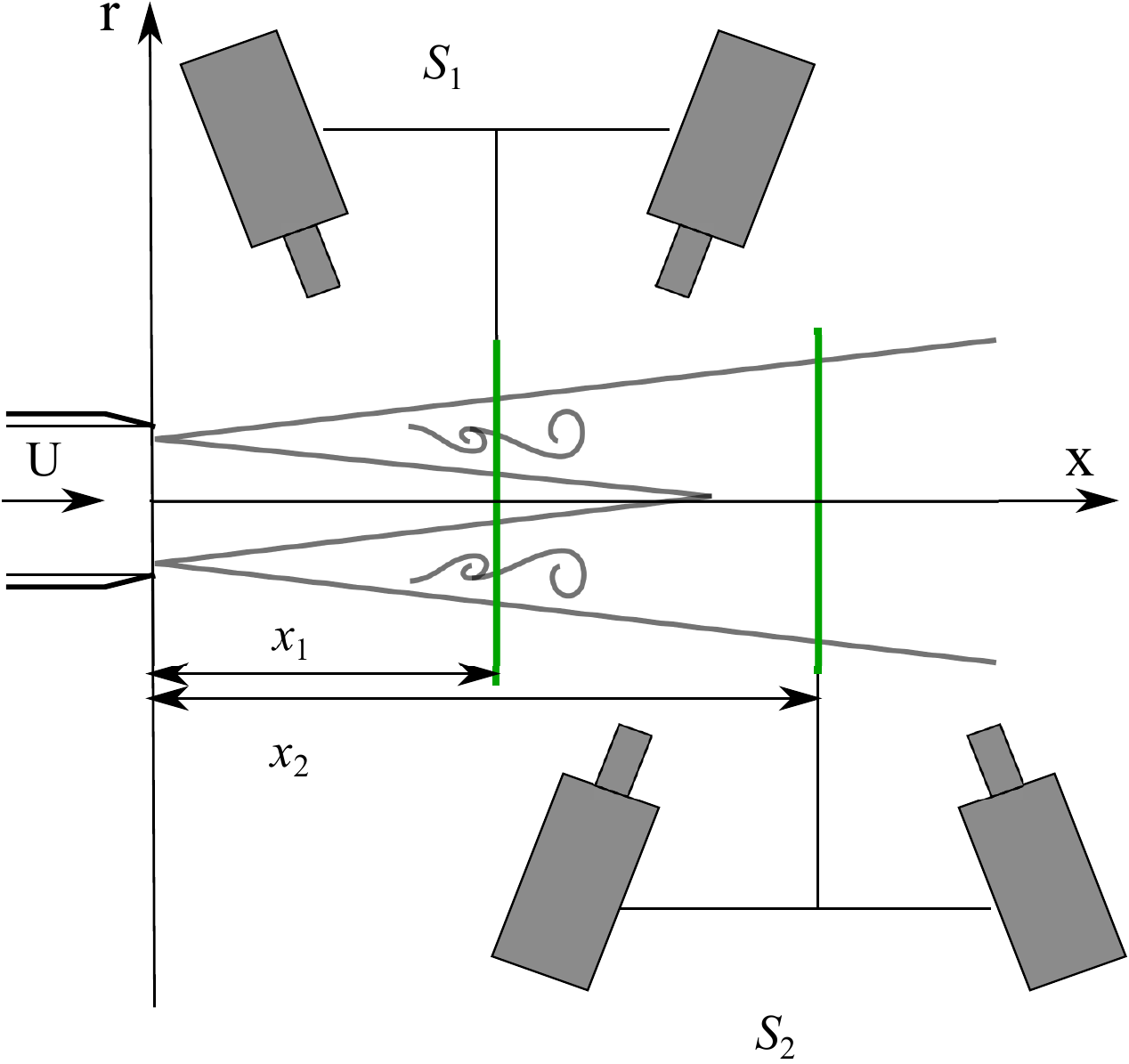}
	\hspace{.5cm}
	\includegraphics[width=0.350\columnwidth]{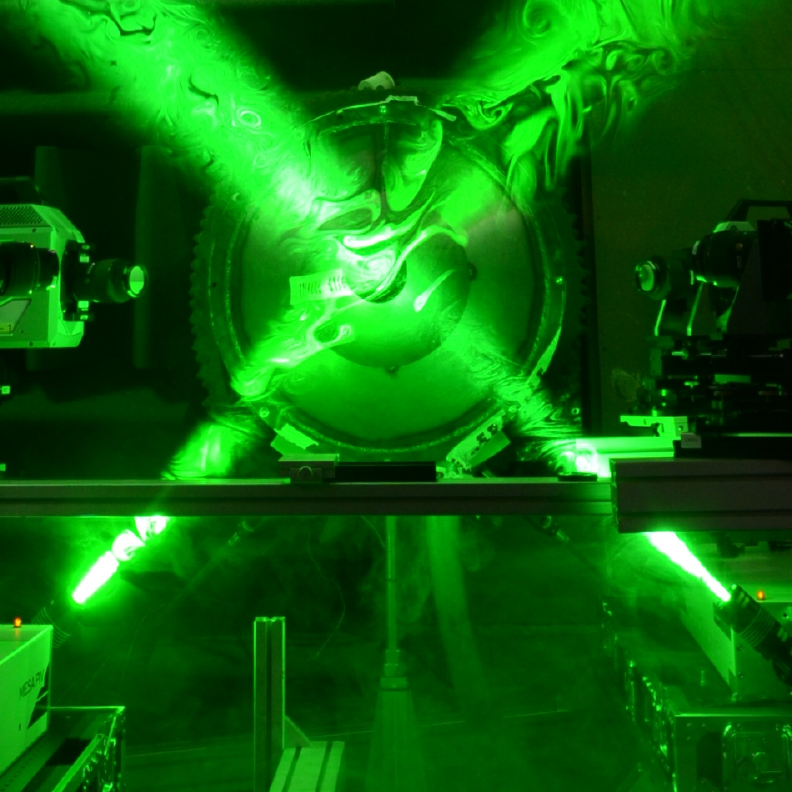}
	\caption{($a$) Sketch of the experimental setup, viewed from the top. The green lines represent the laser light sheets $S_1$ and $S_2$, placed at $x_1$ and $x_2$, respectively; these planes are shifted during the experiment in the range $x=[1, 8]$ (see \cite{jaunet2017prf}). ($b$) Front view of the experiment during the PIV acquisition.}\label{fig:pivsketch}
\end{figure}

The jet experiments of Jaunet \etal~\cite{jaunet2017prf} provide an extensive database of synchronous PIV measurements in cross-planes at several axial positions. Two-point coherence statistics along fixed radial positions have been discussed in the first publication \cite{jaunet2017prf} with a focus on the streamwise coherence length; here, the same database is fully exploited for the computation of the two-dimensional cross-spectral density of axisymmetric velocity fluctuations in the $(x,r)$ plane. To our knowledge, no experimental CSD measurements of comparable size and detail in a turbulent jet exist in the literature.

The acquisition apparatus consists of two time-resolved stereo PIV systems that can be moved independently. Both systems measure the velocity in planes orthogonal to the jet axis at either the same axial location (co-planar configuration) or at different positions. A sketch of the setup is shown in Fig.~\ref{fig:pivsketch}$(a)$, where it is illustrated how the two PIV systems can be positioned with respect to the nozzle. The axial positions of the measurement planes are $x_1 \in [1, 8]$ and $x_2 \in [x_1, 8]$ in increments of $\Delta x = 0.5$, where $x_1$ refers to the axial position of the upstream system ($S_1$) and $x_2$ to that of the downstream system ($S_2$). The instantaneous velocity
fields are interpolated onto a polar grid of 32 points in the radial direction and 64 in azimuth, for $r \le 0.8$, using a bi-cubic interpolation that guarantees a close match with the original data. 

The axisymmetric component of axial velocity fluctuations is isolated by averaging each snapshot in the azimuthal direction. The cross-spectral density (CSD) matrix between all resulting $(N_x\times N_r) = (15 \times 64)$ spatial positions is then constructed using Welch's periodogram method, with data blocks of 128 time samples, overlapped by 50\% (see Ref.~\cite{jaunet2017prf} for further details).  
This empirically constructed matrix converges statistically towards the true CSD, which is defined as the covariance of the Fourier-transformed velocity signal $\hat{y}(\mathbf{x}_i,\omega)$,
\begin{equation}
P_{\hat{y}\hat{y}}\big|_{ij}(\omega) = \mathcal{E}\left[\hat{y}(\mathbf{x}_i,\omega)\hat{y}^*(\mathbf{x}_j,\omega)\right].
\end{equation}
The `expected value' operator $\mathcal{E}$ denotes the asymptotic limit of an ensemble average.
In the present calculations, for numerical reasons, each element of the CSD matrix is further scaled with a factor $\sqrt{r_i r_j}$, composed of the radial coordinates of any two points for which the correlation is computed. This procedure ensures that the resulting modified CSD matrix is strictly Hermitian \cite{citriniti2000reconstruction}.

Eigenvectors $\tilde{\phi}_k$ of the modified matrix are computed. These are then again rescaled in each point as $\phi_k (r,x) = \tilde{\phi}_k(r,x) r^{-0.5}$, and they are sorted in descending order of their associated eigenvalues. The structures $\phi_k (r,x)$ represent the SPOD modes, in the terminology of Picard \& Delville \cite{picard2000pressure}, and as used in recent literature \cite{towne2018,schmidt2018}. Unfortunately, the same name is also used by Sieber \etal~\cite{sieberSPOD} for a different modal decomposition, which is not employed here.

The statistical convergence of SPOD modes is examined by dividing the datasets into two blocks, indicated as $i=(1,2)$, and performing the computation procedure on each subset. Each block corresponds to half of the original dataset. We use a normalised scalar product $\alpha$ between each mode $\phi_{i,k}$ obtained with half of the original dataset and the corresponding mode $\phi_{k}$ obtained with the complete set,
\begin{equation}
\alpha_{i,k} = \frac{\langle \phi_k,\phi_{i,k} \rangle}
{\sqrt{|| \phi_k ||_2 \cdot || \phi_{i,k} ||_2}}.
\label{e:scalar}
\end{equation}
The scalar quantity $\alpha_{i,k}$ is the correlation coefficient between the $k^{th}$ mode of subset $i$ and the corresponding mode of the full dataset. We consider modes with a correlation coefficient close to unity as being converged, showing thus that the same computation with half of the dataset leads to a very similar result.

\begin{figure}	
	\centering
	\includegraphics[width=0.9\textwidth]{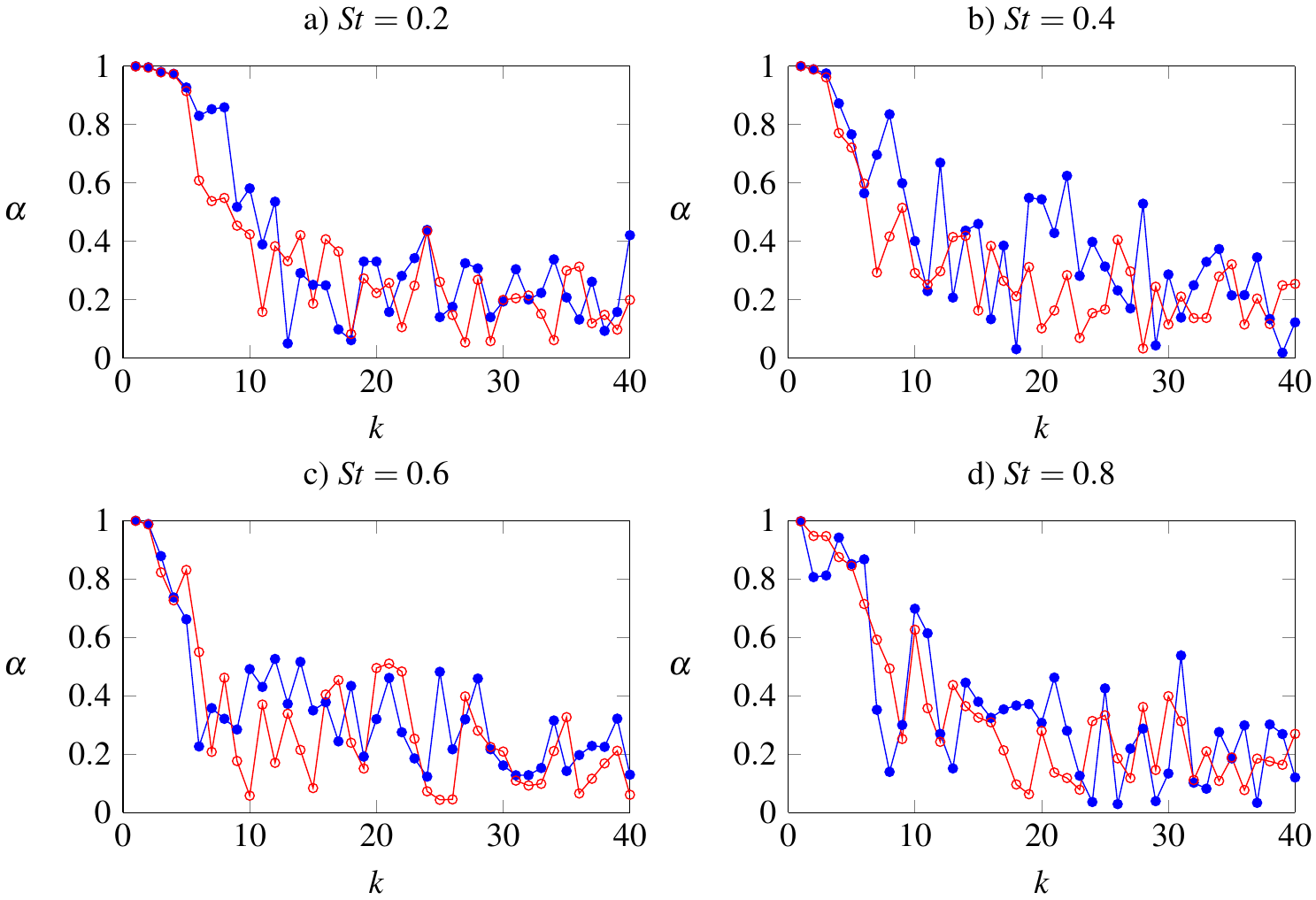}
	\caption{Correlation coefficients $\alpha$ (Eq.~\ref{e:scalar}), for a quantification of the statistical convergence of experimental SPOD modes. Two subsets of data are taken from the experimental acquisitions, and for each of these the SPOD modes are computed and compared: ({\color{blue}$\bullet$}) $\alpha_{1,k}$ and ({\color{red}$\circ$}) $\alpha_{2,k}$. Satisfactory convergence is observed at least for mode 1 (throughout) and for mode 2 (at $St < 0.8$).}
	\label{fig:convergence_CSDPOD}
\end{figure}

The correlation coefficients for $St=0.2,\,0.4,\,0.6$ and 0.8 are presented in Fig.~\ref{fig:convergence_CSDPOD}. It is clear from these figures that the analysis is rather sensitive to the amount of data being used. The discrepancies in the higher (less energetic) modes are partially explained by differences in the order in which they emerge, depending on the data subset. However, the first two SPOD modes seem to be sufficiently correlated and can be accepted as being converged at all Strouhal numbers below $St=0.8$. Only modes 1 and 2 will be discussed in the following, for $0.2\le St \le 0.7$.

\subsection{Resolvent-based modelling of SPOD modes}

The relation between resolvent modes, as presented in Sec.~\ref{sec:resolvent}, and SPOD modes, as obtained from the experiments, is made explicit here on the basis of our earlier formulation \cite{semeraro2016stochastic}. The following development is consistent with the recent work of Towne \etal~\cite{towne2018} and Schmidt \etal~\cite{schmidt2018}, while using the nomenclature introduced in the previous sections.

Let $f(t)$ and $q'(t)$ represent spatial discretisations of the stochastic forcing and response, as discussed in Sec.~\ref{sec:linsystem}. The CSD of their spectral components $\hat{f}(\omega)$ and $\hat{q}(\omega)$ is given by
\begin{equation}\label{eq:csddef}
P_{\!\hat{f}\!\hat{f}}(\omega) = \mathcal{E} \left[\hat{f}(\omega) \hat{f}^H(\omega)\right] \quad \text{and} \quad
P_{\hat{q}\hat{q}}(\omega) = \mathcal{E} \left[\hat{q}(\omega) \hat{q}^H(\omega)\right].
\end{equation}

For the purpose of flow modelling, we consider the CSD of an experimentally observable vector $\hat{y}$ of flow quantities,
\begin{equation}
P_{\hat{y}\hat{y}}(\omega) = \mathcal{E} \left[\hat{y}(\omega) \hat{y}^H(\omega)\right],  \text{~~with~~} \hat{y} = C\hat{q}.
\end{equation}
The relation between $P_{\hat{y}\hat{y}}$ and $P_{\!\hat{f}\!\hat{f}}$ at a given frequency involves the resolvent operator; with (\ref{eqn:resolventdecomp}) and the definitions in Sec.~\ref{sec:optimaz}, this relation can be written as
\begin{equation}
P_{\hat{y}\hat{y}} =
CR\,  \mathcal{E} \left[\hat{f} \hat{f}^H \right] R^H C^H =
CQ\Sigma F^H M P_{\!\hat{f}\!\hat{f}} M F \Sigma Q^H C^H. \label{eq:Pyy}
\end{equation}
If the forcing $\hat{f}$ is expanded in the basis given by the columns of $F$, with a coefficient vector $\beta$ such that $\hat{f}=F\beta$, (\ref{eq:Pyy}) becomes
\begin{equation}
P_{\hat{y}\hat{y}} = CQ\Sigma P_{\beta\!\beta} \Sigma Q^H C^H. \label{eq:PyyAlpha}
\end{equation}
We now seek the relation between eigenvectors (SPOD modes) of $P_{\hat{y}\hat{y}}$ and the resolvent response modes contained in the matrix $Q$. If full-state information is available, $C=I$ and $\hat{y} = \hat{q}$, one can write
\begin{equation}
NP_{\hat{y}\hat{y}}N^H = NQ\Sigma P_{\beta\!\beta} \Sigma Q^H N^H. \label{eq:NPyyN}
\end{equation}
Recall that the scalar product (\ref{eq:norm}) is represented by the matrix $M=N^HN$.
As $NQ$ is unitary, $(NQ)^H NQ=I$, it represents the eigenvector matrix of $NP_{\hat{y}\hat{y}}N^H$, under the condition that $P_{\beta\!\beta}$ is a diagonal matrix. This condition signifies that the resolvent forcing modes, for a given frequency, are \emph{uncorrelated} in the actual stochastic forcing of the system (`spatial white-noise hypothesis'). It finally follows that $Q$ in this case is the eigenvector matrix of $P_{\hat{y}\hat{y}}M$ (the `weighted CSD' \cite{schmidt2018}), with the diagonal elements of $P_{\beta\!\beta}\Sigma^2$ as associated eigenvalues.

If $\hat{y}$ represents only partial-state information, $C\neq I$, such a direct link between resolvent response modes and SPOD modes cannot be made. This is the case for the present experimental dataset. However, under the strong hypothesis $P_{\beta\!\beta}=I$, it is possible to construct a low-rank approximation
\begin{equation}
P_{\hat{y}\hat{y}} \approx C\tilde{Q}\tilde{\Sigma}^2 \tilde{Q}^H C^H \label{eq:Pyy_lowrank}
\end{equation}
of the observable CSD, where $\tilde{Q}$ and $\tilde{\Sigma}$ only contain a limited number of resolvent response modes and associated gains, as obtained from the linear analysis. The eigenvectors of (\ref{eq:Pyy_lowrank}), or equivalently the left singular vectors of $C\tilde{Q}\tilde{\Sigma}$, can then be identified and compared to those computed from the experimental data. It may be expected that the leading SPOD mode structure is well represented by such a linear model in situations where the first optimal gain $\sigma_1$ is significantly larger than $\sigma_2$: as the ratio $\sigma_1/\sigma_2$ tends towards infinity, the leading SPOD mode tends towards the optimal resolvent response mode $C\hat{q}_1$. At finite gain ratios however, the inclusion of several resolvent modes in $C\tilde{Q}\tilde{\Sigma}$ has the potential to improve the agreement. The comparisons provided in recent analyses of backward-facing step flow \cite{beneddine2016conditions} and jets \cite{towne2018,schmidt2018} show such favourable cases of strong gain separation, as discussed in those articles.

\subsection{Comparison between experimental and model results}
\label{sec:comparison}

It is now assessed to what extent the experimentally obtained SPOD modes are accurately reproduced by the resolvent-based model. The success of this comparison depends on many factors, namely, the assumption that our forcing modes are uncorrelated in $P_{\hat{y}\hat{y}}$, the hypotheses involved in the linear resolvent analysis in Sec.~\ref{sec:resolvent}, and the accuracy of both experimental and numerical methods used.

Approximations of $P_{\hat{y}\hat{y}}$ are constructed according to the low-rank model (\ref{eq:Pyy_lowrank}). The first five resolvent modes, discussed in Sec.~\ref{sec:numresults}, are used to build $\tilde{Q}$ and $\tilde{\Sigma}$ (the low-rank versions of $Q$ and $\Sigma$) at various Strouhal numbers. The matrix $C$ selects the streamwise velocity component in the same grid points that are used in the experimental CSDs. SPOD modes are then computed as the left singular modes of the matrix $C\tilde{Q}\tilde{\Sigma}$.
%
%
\begin{figure}
		\includegraphics[width=\textwidth]{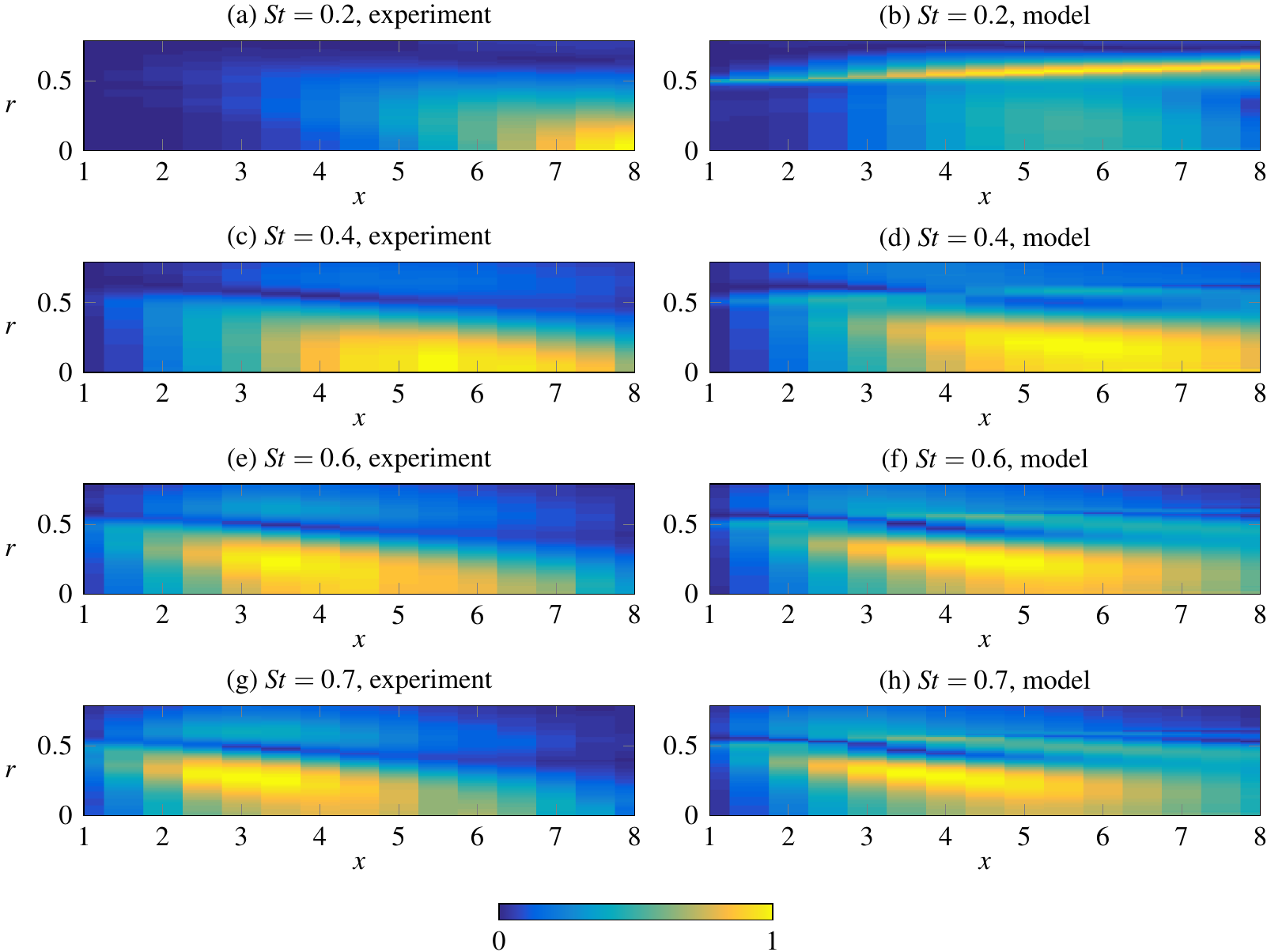}
	\caption{Modulus of the \emph{first} SPOD mode at $St=0.2,\,0.4,\,0.6$ and 0.7, as obtained from the experimental data (left column) and from the resolvent-based model (right column). The resolution of the colour plots corresponds to the spatial grid where the CSD is defined, without interpolation.\label{fig:contou_opt}
	}
\end{figure}
%
%

%
%
\begin{figure}
	\includegraphics[width=\textwidth]{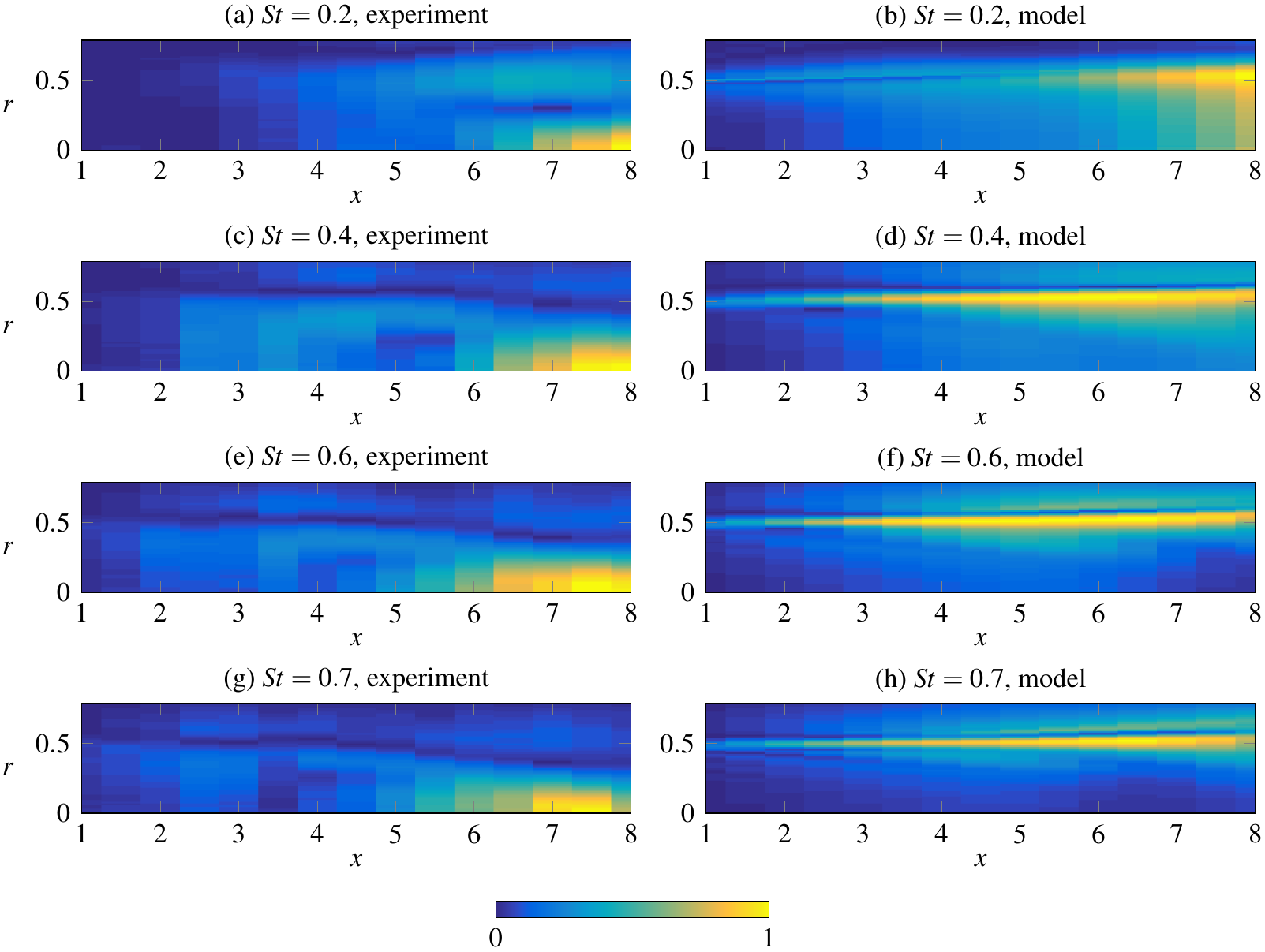}
	\caption{Modulus of the \emph{second} SPOD mode at $St=0.2,\,0.4,\,0.6$ and 0.7, as obtained from the experimental data (left column) and from the resolvent-based model (right column). The resolution of the colour plots corresponds to the spatial grid where the CSD is defined, without interpolation.
		\label{fig:contou_sub}
	}
\end{figure}

The leading SPOD modes obtained from experimental data and from the resolvent-based model are compared in Fig.~\ref{fig:contou_opt}, for Strouhal numbers $St = 0.2$, 0.4, 0.6 and 0.7. Contours of their absolute value are shown, and each mode is normalised with respect to its global maximum value. The agreement between experimental (left column) and model results (right column) is remarkably good at Strouhal numbers between 0.4 and 0.7. Within this range of $St$, maximum SPOD amplitudes are located inside the potential core region of the jet. The maximum along $r$ at each streamwise station follows a line that tends towards the jet axis, evocative of the `critical layer' as discussed by Tissot \etal~\cite{tissot2016sensitivity}. At $St=0.2$ however, the agreement between experiment and model is rather poor. While the experimental mode structure in Fig.~\ref{fig:contou_opt}$a$ resembles those found at higher Strouhal numbers, but with its maximum further downstream and possibly outside the measurement window, the resolvent-based model predicts high amplitudes in the outer portion of the shear layer (Figs.~\ref{fig:contou_opt}$b$ and \ref{fig:real_r_opt}$a$).

The second SPOD modes are shown in the same manner in Fig.~\ref{fig:contou_sub}. For these modes, the comparison between experimental and model results fails at all Strouhal numbers. Mode structures obtained from the resolvent-based model have high amplitudes inside the shear layer, similar to the sub-optimal response structures shown in Fig.~\ref{fig:real_fr_St07}, whereas the experimentally educed structures are still characterised by maximum amplitudes near the jet axis. Inside the jet, the latter display an amplitude modulation along $x$ with two distinct local maxima. Subsequent SPOD modes show similarly poor agreement, and they are not reported here.

Several effects may contribute to the failure of the model to capture the second SPOD mode; 
a rather obvious one seems to derive from the specific structure of the sub-optimal resolvent modes that are included in the low-rank operator (\ref{eq:Pyy_lowrank}). The optimal resolvent mode cannot be significantly involved in the second SPOD mode, which is orthogonal to the first one, and none of the four sub-optimal structures in Fig.~\ref{fig:real_fr_St07} can be expected to reproduce spatial variations of the kind observed in the left column of Fig.~\ref{fig:contou_sub}. 

\begin{figure}
	\centering
	\begin{tabular}{@{} c @{} c @{} c @{}}
		\includegraphics[width=0.45\textwidth]{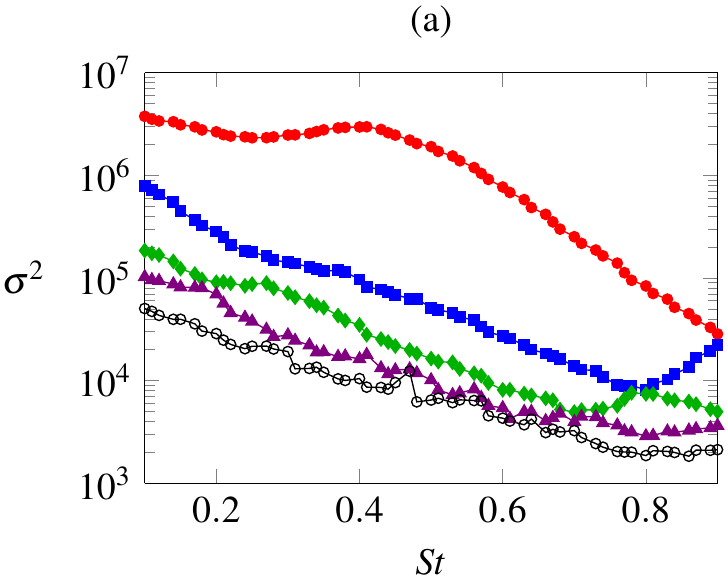}& \hspace{0.05\textwidth} &
		\includegraphics[width=0.45\textwidth]{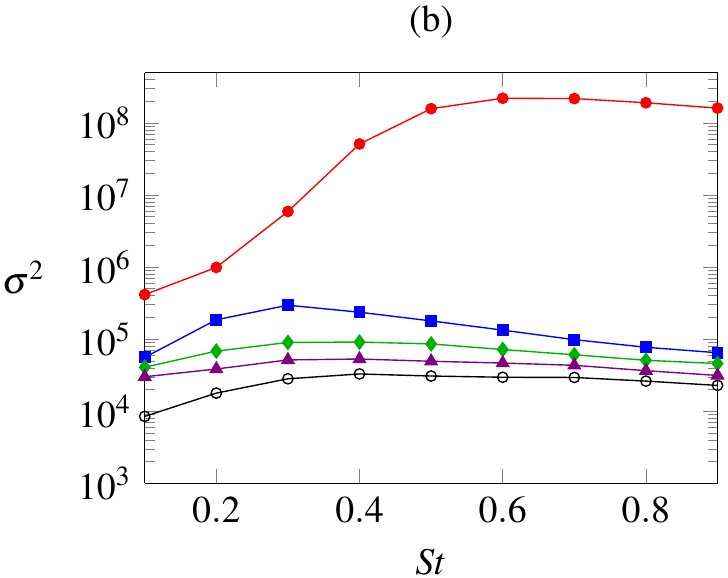}
	\end{tabular}
	\caption{The leading five CSD eigenvalue branches as functions of Strouhal number, (a) from the experiment, (b) from the resolvent-based linear model.}\label{fig:eigen}
\end{figure}

Energy spectra, as given by the eigenvalues of the measured and modelled CSD matrices, are compared in Fig.~\ref{fig:eigen}. Their variations with Strouhal number are quite different from one another. The dominant eigenvalue of the experimental CSD takes on its highest value at $St=0.1$, and another local maximum arises at $St=0.4$. The first and second eigenvalue curves are separated by a factor between 3 and 7 over the interval $0.4\le St \le 0.8$, where SPOD modes in model and experiment are in good agreement. CSD eigenvalues derived from the model closely resemble the gain values shown in Fig.~\ref{fig:gains}, with a slight shift of the maximum value from $St=0.7$ to 0.6. An important source of discrepancy between the dominant branches in figures \ref{fig:eigen}$a$ and $b$ is very likely the assumption that all resolvent forcing modes over all Strouhal numbers are contained in the Reynolds stress fluctuations with \emph{equal amplitude}.
\begin{figure}
		\includegraphics[width=\textwidth]{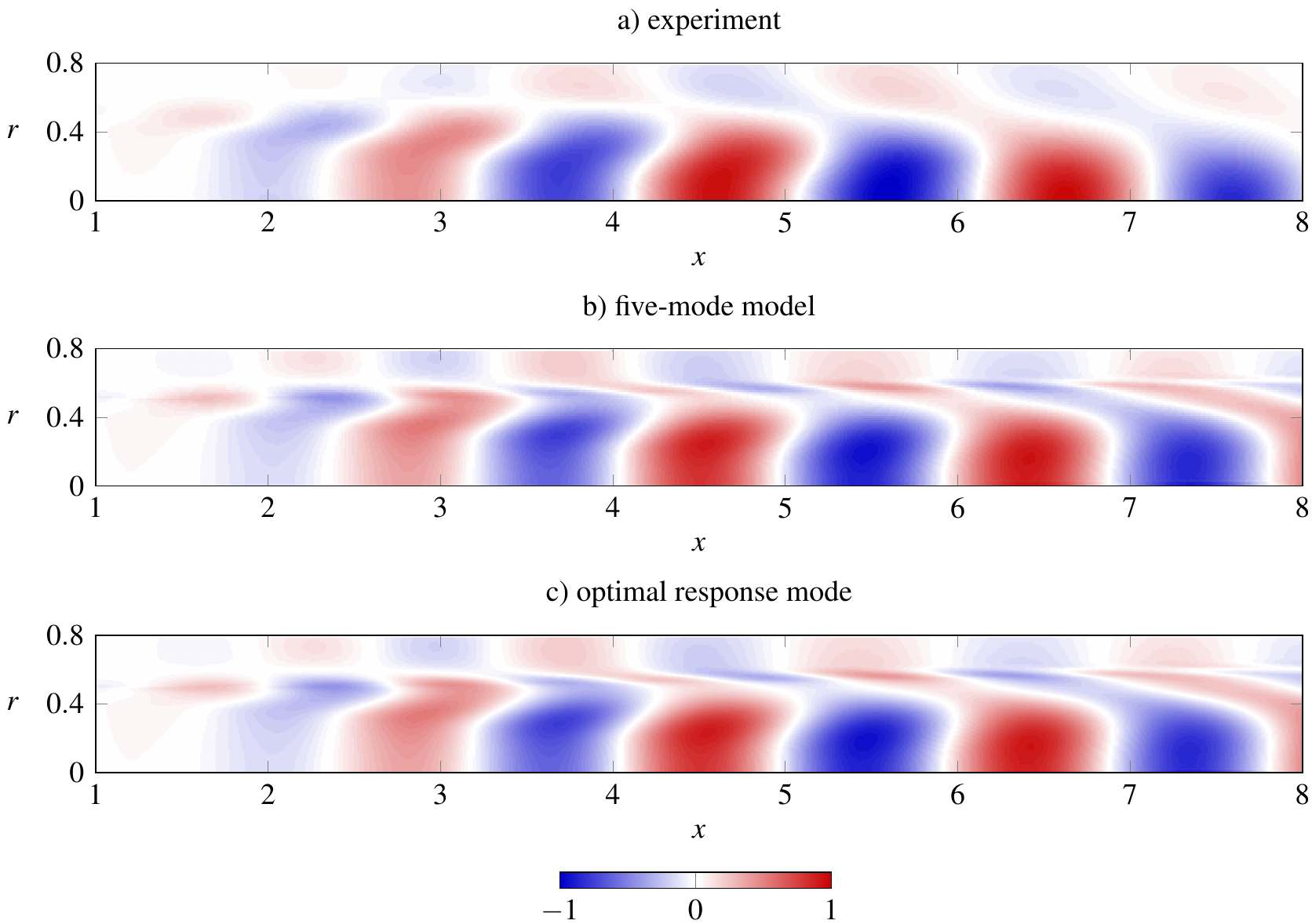}
	\caption{Interpolated SPOD wavepackets at $St=0.4$, ($a$) from the experiment, ($b$) from the linear model based on five resolvent modes. Amplitude and phase are interpolated between the available data points, and the resulting real part is represented. ($c$) The corresponding \emph{optimal} response mode alone. \label{fig:real_opt_St04}}
\end{figure}

In order to visualise the wavepacket structure of the leading SPOD mode at $St=0.4$, the dominant $St$ value according to the experimental spectrum (Fig.~\ref{fig:eigen}a),
the amplitude and the phase of this mode are interpolated onto a fine mesh. This is done both for the experimental and for the model SPOD mode, and the resulting real parts are shown in Fig.~\ref{fig:real_opt_St04}$a,b$. Clean wavepackets are recovered, and their resemblance is even more appreciable than in the amplitude plots of Fig.~\ref{fig:contou_opt}. Recall that the model SPOD mode has been obtained as the eigenmode of a CSD matrix that was constructed from the first five resolvent response modes at $St=0.4$. The optimal response mode alone is plotted in Fig.~\ref{fig:real_opt_St04}$c$; although not strictly identical, it is indeed virtually indistinguishable from the five-mode model result. This comparison demonstrates that the extra effort of including sub-optimal response modes in the model has not led to any improvement of the SPOD prediction: the `rank 1 approximation' used in previous studies \cite{beneddine2016conditions,schmidt2018}, which consists in equating the optimal response with the first SPOD mode, is applicable in the current jet case, and sub-optimal modes are of no use for increasing the accuracy of the linear model.

\begin{figure}
		\includegraphics[width=\textwidth]{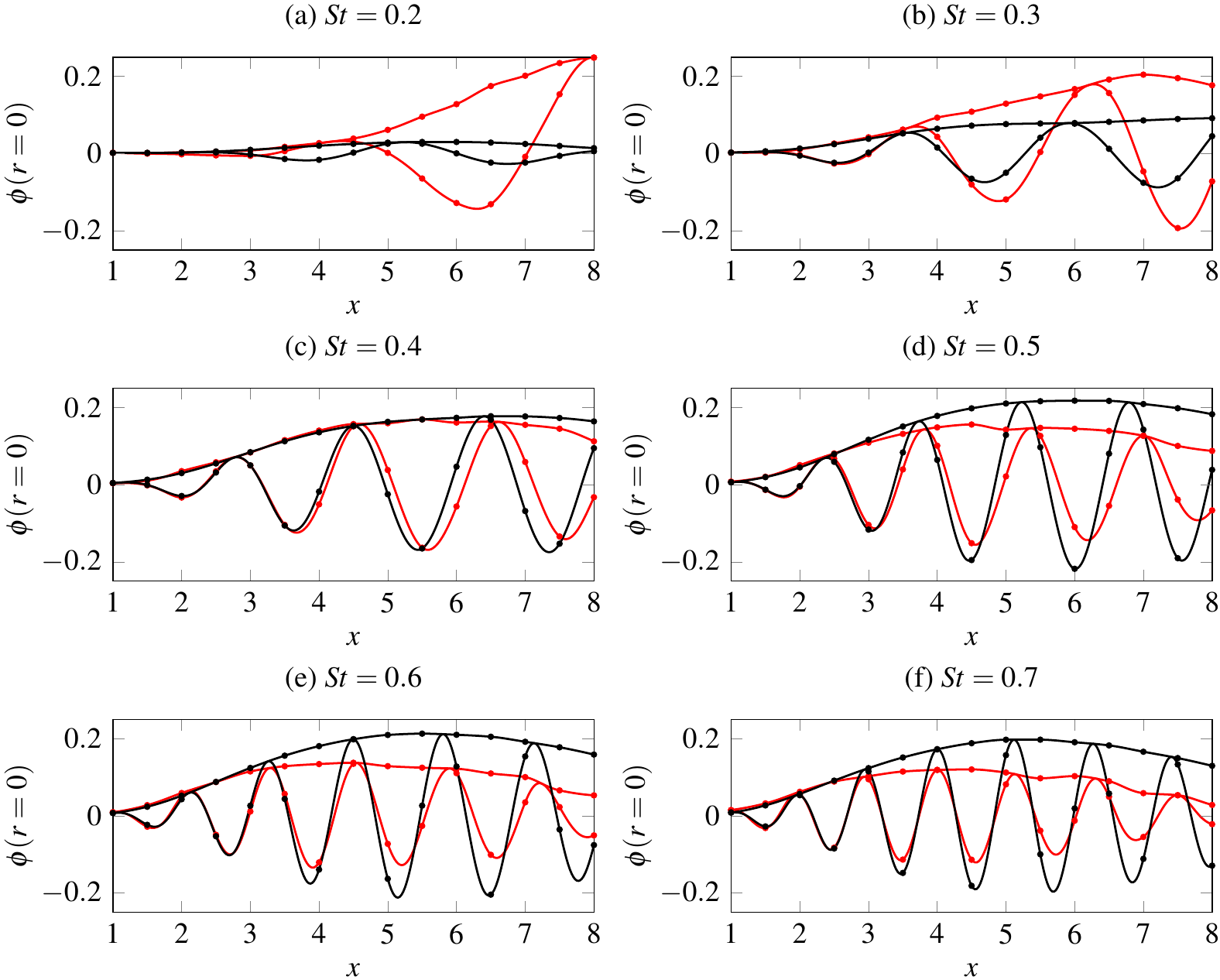}
	\caption{Comparison between experimental (red) and model (black) SPOD wavepackets at various Strouhal numbers. Amplitude and phase variations on the centreline are interpolated between the available data points, indicated by markers.  Both the amplitude envelopes and the oscillating real parts are shown.\label{fig:wave_1}
	}
\end{figure}

For a more quantitative comparison, real-part oscillations of the interpolated SPOD modes are extracted along the centreline, and displayed in Fig.~\ref{fig:wave_1}, together with their amplitude envelope. Black and red lines represent the five-mode model and the experimental data, respectively. Markers indicate the values obtained directly in the original measurement points. Experimental wavepackets are traced with their actual absolute amplitude, whereas a best-fit coefficient has been constructed, based on the interval $1\le x \le 5$, for a proper scaling of the model amplitude. Good agreement is generally observed in the upstream region of exponential amplitude growth; at $St=0.4$, the agreement is excellent down to the amplitude maximum. At lower Strouhal numbers, the model underpredicts the maximum, even by a large measure in the case of $St=0.2$, whereas at higher Strouhal numbers, the amplitude maximum is overpredicted. Considering that a phase match is imposed in the very first position, $x=1$, and differences therefore accumulate in the downstream direction, the phase prediction can be said to be satisfactory for all Strouhal numbers above $0.2$. Several radial positions have been tested for the present comparison, and all have been found to give very similar agreement. The most notable difference between model and experimental results, at $St\ge 0.4$, is an underpredicted downstream attenuation of fluctuation amplitudes. This trend is clearly visible in Figs.~\ref{fig:contou_opt}, \ref{fig:real_opt_St04} and \ref{fig:wave_1}, and it increases with $St$. 

It must be kept in mind that individual wavepackets in the present approach are regarded as isolated objects, which is made possible by our choice to replace the nonlinear term with a generic white noise forcing.  In reality, all frequencies and azimuthal wavenumbers are coupled through the nonlinear Reynolds stresses, such that energy is exchanged between coherent structures, axisymmetric and non-axisymmetric, at different frequencies. This energy transfer would be correctly represented by spatial variations of the forcing at a given Strouhal number, which influence the wavepacket envelope, while furthermore the forcing distributions at all Strouhal numbers are coupled among each other. It remains a challenge for future work to identify a consistent way to model these interactions in a turbulent flow with a broadband spectrum of frequencies and azimuthal wavenumbers.
%
%

\subsection{Comparison with linear jet studies in the recent literature}

The resolvent modes presented in Sec.~\ref{sec:numresults} and their comparison with SPOD modes in Sec.~\ref{sec:comparison} are, by and large, consistent with the findings of similar recent studies \cite{garnaud2013preferred,jeun2016input,semeraro2016modeling,towne2018,schmidt2018}. One striking difference with the results of Schmidt \etal~\cite{schmidt2018} is noted in the structure of sub-optimal response modes: at Strouhal numbers above 0.2, our computations yield a clean separation between modal shear and non-modal Orr structures, whereas Schmidt \etal~\cite{schmidt2018} observe a mixing of shear-induced wavepackets with Orr-related structures in all their sub-optimals. The associated forcing structures, shown in Figs.~\ref{fig:real_f_opt} and \ref{fig:real_fr_St07}, suggest that this difference can be attributed to the presence of a nozzle in our numerical configuration. Forcing inside the pipe is found to be particularly efficient, especially in the case of the optimal resolvent mode, which must therefore be expected to be very sensitive to the truncation of the most receptive flow region. In turn, changes in the optimal mode will be accompanied by changes in the orthogonal sub-optimals. The localisation of optimal forcing in the present results may furthermore be linked to the observed sensitivity of LES statistics with respect to flow details in the nozzle boundary layer \cite{bres2018importance}. 
%
%
\begin{figure}
	\includegraphics[width=\textwidth]{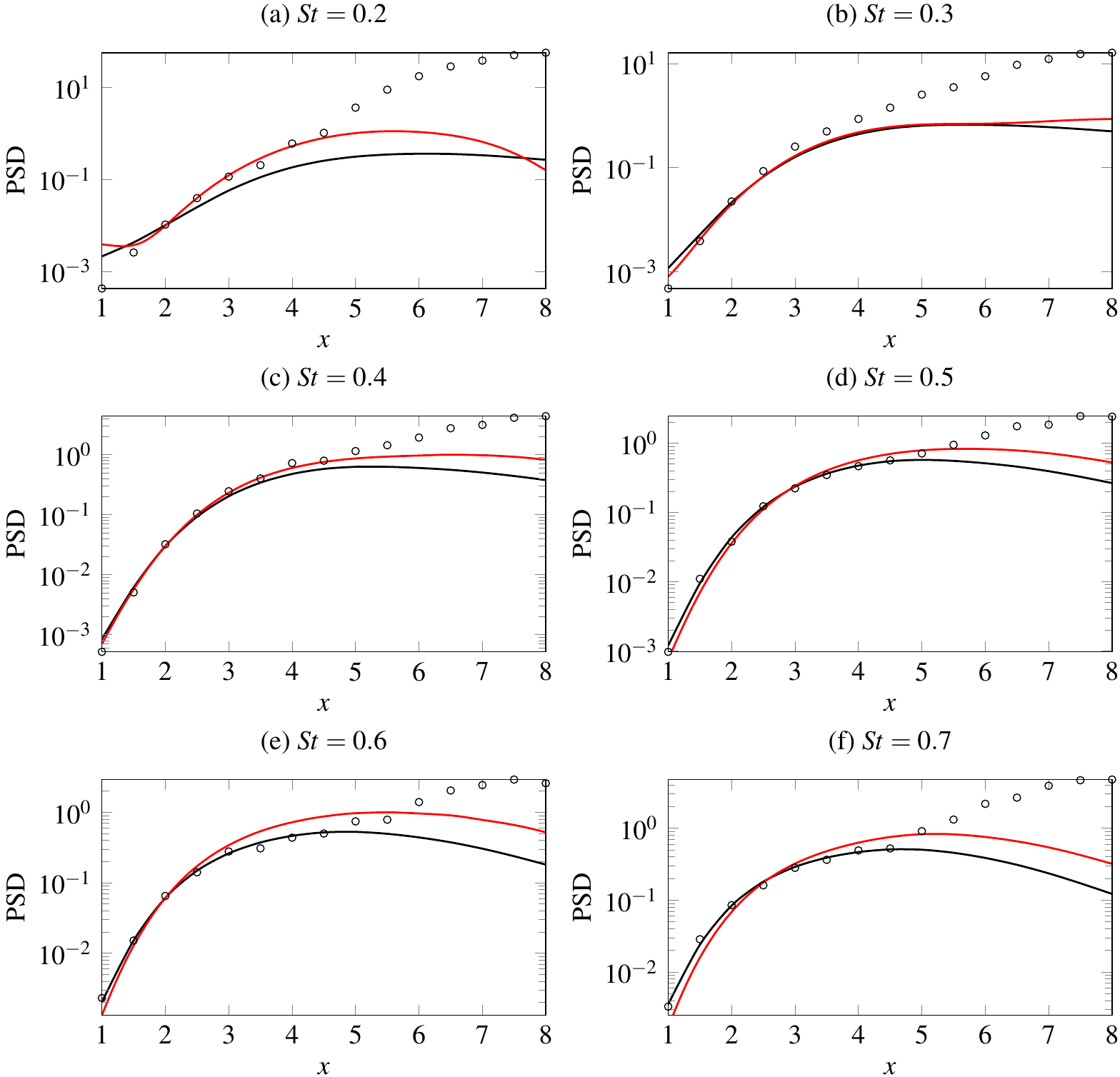}
	\caption{Power-spectral density (PSD) along the jet centreline as a function of $x$ at different Strouhal numbers. ($\circ$) Hot-wire measurements \cite{cavalieri2013wavepackets}; (\rule[2pt]{10pt}{1pt}) PSE model \cite{cavalieri2013wavepackets}; (\textcolor{red}{\rule[2pt]{10pt}{1pt}}) optimal response mode (present calculations).\label{fig:PSD}}
\end{figure}

Numerous previous studies addressing the linear modelling of wavepackets in turbulent jets, when only \emph{boundary forcing} at the inflow was considered, observed discrepancies in the initial amplitude growth at low Strouhal numbers, typically $St \leq 0.3$ \cite{gudmundsson2011instability,cavalieri2013wavepackets,breakey2013near}, which play an important role in the generation of jet noise. These differences were initially attributed either to the weakly non-parallel assumption that underlies the parabolised stability equations (PSE), or to unmodelled nonlinear effects. The limitation of PSE with regard to non-parallelism has since been ruled out by computations based on the fully non-parallel, linearised Euler equations \cite{baqui2015coherence}. We complete this study by revisiting the low-Strouhal discrepancy in the resolvent framework, where \emph{volume forcing} is included as a surrogate for nonlinear effects.

Figure \ref{fig:PSD} compares the PSE and measured power-spectral density (PSD) results from Cavalieri \etal~\cite{cavalieri2013wavepackets} with the kinetic energy of the optimal response modes presented in Sec.~\ref{sec:numresults}. All curves are extracted on the jet axis. It can be seen how, at $St=0.2$, the resolvent mode captures the initial wavepacket growth measured in the experiment, and underpredicted by PSE, while at higher Strouhal numbers the PSE solution and resolvent mode comprise similar initial growth. The optimal forcing at $St=0.2$ (Fig.~\ref{fig:real_f_opt}a,b) involves significant contributions from inside the shear layer of the free jet, which were excluded in the models of Refs.~\cite{gudmundsson2011instability,cavalieri2013wavepackets,breakey2013near,baqui2015coherence}. This volume forcing boosts the spatial growth of the response wavepacket near the nozzle. Although the high-amplitude portion of the optimal response mode is not in agreement with the experimental PSD, the initial growth is faithfully reproduced. This observation suggests that Orr-type forcing through Reynolds stresses in the free shear layer indeed contributes to perturbation growth at low Strouhal number near the nozzle.

Another much-discussed discrepancy between PSE and PSD results arises in the downstream region where linear models predict a decay in perturbation amplitude, whereas the PSD displays a marked additional growth, accompanied by a slope break in the growth rate. Jordan \etal~\cite{jordan2017modal} interpret the PSD growth in this region in terms of the non-normality of decaying local eigenmodes. However, this peculiar behaviour of the PSD is entirely absent in the present SPOD results, in agreement with the LES-based analysis by Schmidt \etal~\cite{schmidt2018}. This leads us to conclude that the spatio-temporal coherence, from which SPOD modes are derived, provides a sharper and more pertinent criterion for the eduction of coherent turbulent structures than the PSD, which only measures temporal coherence. As discussed by Towne \etal~\cite{towne2018}, the PSD may contain the trace of many SPOD modes. Figure \ref{fig:PSD} strongly suggests that sub-optimal SPOD or resolvent mode wavepackets contribute in a very significant way to the turbulent dynamics downstream of $x=4$. 
The Orr-type character of our sub-optimal resolvent modes is fully consistent with the discussion of the PSD behaviour by Tissot \etal~\cite{tissot2016sensitivity}.


\section{Conclusions}\label{sec:conclusion}

Perturbation wavepackets in the mean flow of a turbulent jet have been computed in the form of resolvent modes. Nonlinear terms in the governing equations, which arise in the form of generalised Reynolds stresses, are regarded as generic forcing terms \cite{farrell1996generalized,mckeon2010critical,beneddine2016conditions}. The five leading orthogonal forcing/response modes have been identified for several values of the Strouhal number between 0.2 and 1.5. The most amplified (`optimal') mode, over the interval $0.3 \le St \le 1.5$, bears the traits of a shear instability in the free jet, with strong spatial growth in the potential core. This mode arises principally from forcing in the nozzle boundary layer, which takes the shape of tilted structures indicative of the Orr mechanism, as described in our earlier studies \cite{garnaud2013preferred,semeraro2016modeling}. The dominant role of shear instability in the free-jet portion of the optimal resolvent mode has been demonstrated by projecting the response wavepacket onto a local $k^+$ shear instability mode. Subsequent (`sub-optimal') modes, with significantly lower energy gain, appear to exploit the Orr mechanism in the free jet. This interpretation is supported via an analogy with optimal forcing in a fully developed parallel jet. At low Strouhal number, $St=0.2$, both shear and Orr mechanisms in the free jet seem to contribute to the optimal resolvent mode in a mixed fashion.

Coherent structures have then been extracted from experimental measurements, in the form of eigenvectors of the cross-spectral density, named `spectral POD' (SPOD) modes \cite{picard2000pressure,towne2018}. Following recent works \cite{semeraro2016stochastic,towne2018}, it has been demonstrated that such modes should, in theory, correspond to the optimal response mode described above, under two strong conditions: (i) the corresponding optimal forcing modes are statistically \emph{uncorrelated} among each other in the nonlinear dynamics, which are interpreted in the linear model as forcing terms;  (ii) the SPOD modes are extracted from \emph{full-state} information. As full-state information is not available from the experimental dataset, the five leading response modes were instead used to construct a low-rank model of the cross-spectral density, under the even stronger assumption that the corresponding optimal forcing structures are uncorrelated and of \emph{equal amplitude} in the nonlinear dynamics. This procedure constitutes our resolvent-based linear model for the statistical dynamics of coherent turbulence structures, as characterised by two-point covariance. 

Very good agreement has been found between the leading SPOD modes as obtained from the experiment and from the resolvent-based model, in a range of Strouhal numbers around 0.4. The leading SPOD mode of the linear model is in fact nearly identical to the optimal response wavepacket, such that the intermediate step of building a model CSD from several response structures turned out to be unimportant for the comparison with the experiment. At $St=0.4$, the model reproduces accurately both the amplitude variations over three decades, down to at least 7 diameters behind the nozzle, and the phase variations in the extracted educed SPOD mode. The maximum wavepacket amplitude is underpredicted at $St<0.4$, and overpredicted at $St>0.4$. At all Strouhal numbers between 0.2 and 0.7, the initial streamwise perturbation growth close to the nozzle is very well retrieved. 

Subsequent (non-leading) SPOD modes of the experimental data and the linear model do not show satisfactory agreement. The discussion of their discrepancies may be approached from two ends: on the one hand, the linear model probably cannot replicate the experimental results because of the restricted number of basis vectors, and because the above-mentioned modelling hypotheses are too crude in order to reproduce the dynamics beyond leading order. On the other hand, the experimental measurements may be too sparse, particularly in terms of spatial resolution, in order to detect the rather fine-scale structures that the linear model predicts.

In summary, the results presented in this paper demonstrate that linear resolvent analysis, performed around the spatially developing, time-averaged mean flow, represents a valid tool for the modelling of coherent wavepacket structures in a stochastically driven turbulent jet. Only the mean flow is required for the construction of this linear model. Wavepackets arising from shear instability, which experience the strongest energy gain, could be matched between model and experiment at Strouhal numbers between 0.2 and 0.7. While these general conclusions corroborate those of the parallel study by Schmidt \etal~\cite{schmidt2018}, performed on the basis of LES data for the same flow configuration, differences are observed in the resolvent mode structures. These relate to the separation of shear and Orr mechanisms in the optimal and sub-optimal modes, and they are attributed to the inclusion of a nozzle in the present analysis.

From a final comparison with earlier PSD measurements \cite{cavalieri2013wavepackets}, it is inferred that sub-optimal SPOD modes seem to play a determining role near and beyond the end of the potential core region. While the link between these modes and free-jet Orr-type growth mechanisms is one more time predicted by the present analysis, poor agreement is found between sub-optimal structures in model and experiment. Further progress of wavepacket modelling in high Reynolds number turbulent jets requires establishing the dynamics that dominate in the flow region downstream of the potential core, and how best to model them.

\begin{acknowledgments} This work was supported by the Agence Nationale de la Recherche under the \emph{Cool Jazz}  project, grant number ANR-12-BS09-0024, and by the D\'el\'egation G\'en\'erale de l'Armement under grant number 2015.60.0004.00.470.75.01. Stability calculations were performed using HPC resources of TGCC and CINES under the allocation x2016-2a6451 made by GENCI.
\end{acknowledgments}


%

\begin{thebibliography}{59}%
	\makeatletter
	\providecommand \@ifxundefined [1]{%
		\@ifx{#1\undefined}
	}%
	\providecommand \@ifnum [1]{%
		\ifnum #1\expandafter \@firstoftwo
		\else \expandafter \@secondoftwo
		\fi
	}%
	\providecommand \@ifx [1]{%
		\ifx #1\expandafter \@firstoftwo
		\else \expandafter \@secondoftwo
		\fi
	}%
	\providecommand \natexlab [1]{#1}%
	\providecommand \enquote  [1]{``#1''}%
	\providecommand \bibnamefont  [1]{#1}%
	\providecommand \bibfnamefont [1]{#1}%
	\providecommand \citenamefont [1]{#1}%
	\providecommand \href@noop [0]{\@secondoftwo}%
	\providecommand \href [0]{\begingroup \@sanitize@url \@href}%
	\providecommand \@href[1]{\@@startlink{#1}\@@href}%
	\providecommand \@@href[1]{\endgroup#1\@@endlink}%
	\providecommand \@sanitize@url [0]{\catcode `\\12\catcode `\$12\catcode
		`\&12\catcode `\#12\catcode `\^12\catcode `\_12\catcode `\%12\relax}%
	\providecommand \@@startlink[1]{}%
	\providecommand \@@endlink[0]{}%
	\providecommand \url  [0]{\begingroup\@sanitize@url \@url }%
	\providecommand \@url [1]{\endgroup\@href {#1}{\urlprefix }}%
	\providecommand \urlprefix  [0]{URL }%
	\providecommand \Eprint [0]{\href }%
	\providecommand \doibase [0]{http://dx.doi.org/}%
	\providecommand \selectlanguage [0]{\@gobble}%
	\providecommand \bibinfo  [0]{\@secondoftwo}%
	\providecommand \bibfield  [0]{\@secondoftwo}%
	\providecommand \translation [1]{[#1]}%
	\providecommand \BibitemOpen [0]{}%
	\providecommand \bibitemStop [0]{}%
	\providecommand \bibitemNoStop [0]{.\EOS\space}%
	\providecommand \EOS [0]{\spacefactor3000\relax}%
	\providecommand \BibitemShut  [1]{\csname bibitem#1\endcsname}%
	\let\auto@bib@innerbib\@empty
	\bibitem [{\citenamefont {Jordan}\ and\ \citenamefont
		{Colonius}(2013)}]{jordan2013wave}%
	\BibitemOpen
	\bibfield  {author} {\bibinfo {author} {\bibfnamefont {P.}~\bibnamefont
			{Jordan}}\ and\ \bibinfo {author} {\bibfnamefont {T.}~\bibnamefont
			{Colonius}},\ }\bibfield  {title} {\enquote {\bibinfo {title} {Wave packets
				and turbulent jet noise},}\ }\href@noop {} {\bibfield  {journal} {\bibinfo
			{journal} {Annu. Rev. Fluid Mech.}\ }\textbf {\bibinfo {volume} {45}},\
		\bibinfo {pages} {173--195} (\bibinfo {year} {2013})}\BibitemShut {NoStop}%
	\bibitem [{\citenamefont {Michalke}(1971)}]{michalke1971instability}%
	\BibitemOpen
	\bibfield  {author} {\bibinfo {author} {\bibfnamefont {A.}~\bibnamefont
			{Michalke}},\ }\bibfield  {title} {\enquote {\bibinfo {title} {Instabilit\"at eines kompressiblen runden {F}reistrahls unter {B}er\"ucksichtigung des {E}influsses der {S}trahlgrenzschichtdicke},}\ }\href@noop {} {\bibfield  {journal}
		{\bibinfo  {journal} {Z. Flugwiss.}\ }\textbf {\bibinfo {volume} {19}},\
		\bibinfo {pages} {319--328} (\bibinfo {year} {1971})}\BibitemShut {NoStop}%
	\bibitem [{\citenamefont {Crighton}\ and\ \citenamefont
		{Gaster}(1976)}]{crighton1976stability}%
	\BibitemOpen
	\bibfield  {author} {\bibinfo {author} {\bibfnamefont {D.~G.}\ \bibnamefont
			{Crighton}}\ and\ \bibinfo {author} {\bibfnamefont {M.}~\bibnamefont
			{Gaster}},\ }\bibfield  {title} {\enquote {\bibinfo {title} {Stability of
				slowly diverging jet flow},}\ }\href@noop {} {\bibfield  {journal} {\bibinfo
			{journal} {‎J. Fluid Mech.}\ }\textbf {\bibinfo {volume} {77}},\ \bibinfo
		{pages} {397--413} (\bibinfo {year} {1976})}\BibitemShut {NoStop}%
	\bibitem [{\citenamefont {Oberleithner}\ \emph {et~al.}(2014)\citenamefont
		{Oberleithner}, \citenamefont {Rukes},\ and\ \citenamefont
		{Soria}}]{oberleithner2014mean}%
	\BibitemOpen
	\bibfield  {author} {\bibinfo {author} {\bibfnamefont {K.}~\bibnamefont
			{Oberleithner}}, \bibinfo {author} {\bibfnamefont {L.}~\bibnamefont {Rukes}},
		\ and\ \bibinfo {author} {\bibfnamefont {J.}~\bibnamefont {Soria}},\
	}\bibfield  {title} {\enquote {\bibinfo {title} {Mean flow stability analysis
				of oscillating jet experiments},}\ }\href@noop {} {\bibfield  {journal}
		{\bibinfo  {journal} {J. Fluid Mech.}\ }\textbf {\bibinfo {volume} {757}},\
		\bibinfo {pages} {1--32} (\bibinfo {year} {2014})}\BibitemShut {NoStop}%
	\bibitem [{\citenamefont {Gudmundsson}\ and\ \citenamefont
		{Colonius}(2011)}]{gudmundsson2011instability}%
	\BibitemOpen
	\bibfield  {author} {\bibinfo {author} {\bibfnamefont {K.}~\bibnamefont
			{Gudmundsson}}\ and\ \bibinfo {author} {\bibfnamefont {T.}~\bibnamefont
			{Colonius}},\ }\bibfield  {title} {\enquote {\bibinfo {title} {Instability
				wave models for the near-field fluctuations of turbulent jets},}\ }\href@noop
	{} {\bibfield  {journal} {\bibinfo  {journal} {‎J. Fluid Mech.}\ }\textbf
		{\bibinfo {volume} {689}},\ \bibinfo {pages} {97--128} (\bibinfo {year}
		{2011})}\BibitemShut {NoStop}%
	\bibitem [{\citenamefont {Baqui}\ \emph {et~al.}(2015)\citenamefont {Baqui},
		\citenamefont {Agarwal}, \citenamefont {Cavalieri},\ and\ \citenamefont
		{Sinayoko}}]{baqui2015coherence}%
	\BibitemOpen
	\bibfield  {author} {\bibinfo {author} {\bibfnamefont {Y.~B.}\ \bibnamefont
			{Baqui}}, \bibinfo {author} {\bibfnamefont {A.}~\bibnamefont {Agarwal}},
		\bibinfo {author} {\bibfnamefont {A.~V.~G.}\ \bibnamefont {Cavalieri}}, \
		and\ \bibinfo {author} {\bibfnamefont {S.}~\bibnamefont {Sinayoko}},\
	}\bibfield  {title} {\enquote {\bibinfo {title} {A coherence-matched linear
				source mechanism for subsonic jet noise},}\ }\href@noop {} {\bibfield
		{journal} {\bibinfo  {journal} {‎J. Fluid Mech.}\ }\textbf {\bibinfo
			{volume} {776}},\ \bibinfo {pages} {235--267} (\bibinfo {year}
		{2015})}\BibitemShut {NoStop}%
	\bibitem [{\citenamefont {Farrell}\ and\ \citenamefont
		{Ioannou}(1993)}]{farrell1993stochastic}%
	\BibitemOpen
	\bibfield  {author} {\bibinfo {author} {\bibfnamefont {B.~F.}\ \bibnamefont
			{Farrell}}\ and\ \bibinfo {author} {\bibfnamefont {P.~J.}\ \bibnamefont
			{Ioannou}},\ }\bibfield  {title} {\enquote {\bibinfo {title} {Stochastic
				forcing of the linearized {N}avier--{S}tokes equations},}\ }\href@noop {}
	{\bibfield  {journal} {\bibinfo  {journal} {Phys. Fluids A}\ }\textbf
		{\bibinfo {volume} {5}},\ \bibinfo {pages} {2600--2609} (\bibinfo {year}
		{1993})}\BibitemShut {NoStop}%
	\bibitem [{\citenamefont {Farrell}\ and\ \citenamefont
		{Ioannou}(2014)}]{farrell2014statistical}%
	\BibitemOpen
	\bibfield  {author} {\bibinfo {author} {\bibfnamefont {B.~F.}\ \bibnamefont
			{Farrell}}\ and\ \bibinfo {author} {\bibfnamefont {P.~J.}\ \bibnamefont
			{Ioannou}},\ }\bibfield  {title} {\enquote {\bibinfo {title} {Statistical
				state dynamics: a new perspective on turbulence in shear flow},}\ }\href@noop
	{} {\bibfield  {journal} {\bibinfo  {journal} {arXiv preprint 1412.8290}\ }
		(\bibinfo {year} {2014})}\BibitemShut {NoStop}%
	\bibitem [{\citenamefont {Schmid}(2007)}]{schmid2007}%
	\BibitemOpen
	\bibfield  {author} {\bibinfo {author} {\bibfnamefont {P.~J.}\ \bibnamefont
			{Schmid}},\ }\bibfield  {title} {\enquote {\bibinfo {title} {Nonmodal
				stability theory},}\ }\href@noop {} {\bibfield  {journal} {\bibinfo
			{journal} {Annu. Rev. Fluid Mech.}\ }\textbf {\bibinfo {volume} {39}},\
		\bibinfo {pages} {129--162} (\bibinfo {year} {2007})}\BibitemShut {NoStop}%
	\bibitem [{\citenamefont {Bagheri}\ \emph {et~al.}(2009)\citenamefont
		{Bagheri}, \citenamefont {Henningson}, \citenamefont {Hoepffner},\ and\
		\citenamefont {Schmid}}]{bagheri2009input}%
	\BibitemOpen
	\bibfield  {author} {\bibinfo {author} {\bibfnamefont {S.}~\bibnamefont
			{Bagheri}}, \bibinfo {author} {\bibfnamefont {D.~S.}\ \bibnamefont
			{Henningson}}, \bibinfo {author} {\bibfnamefont {J.}~\bibnamefont
			{Hoepffner}}, \ and\ \bibinfo {author} {\bibfnamefont {P.~J.}\ \bibnamefont
			{Schmid}},\ }\bibfield  {title} {\enquote {\bibinfo {title} {Input-output
				analysis and control design applied to a linear model of spatially developing
				flows},}\ }\href@noop {} {\bibfield  {journal} {\bibinfo  {journal} {Appl.
				Mech. Rev.}\ }\textbf {\bibinfo {volume} {62}},\ \bibinfo {pages} {020803}
		(\bibinfo {year} {2009})}\BibitemShut {NoStop}%
	\bibitem [{\citenamefont {Picard}\ and\ \citenamefont
		{Delville}(2000)}]{picard2000pressure}%
	\BibitemOpen
	\bibfield  {author} {\bibinfo {author} {\bibfnamefont {C.}~\bibnamefont
			{Picard}}\ and\ \bibinfo {author} {\bibfnamefont {J.}~\bibnamefont
			{Delville}},\ }\bibfield  {title} {\enquote {\bibinfo {title} {Pressure
				velocity coupling in a subsonic round jet},}\ }\href@noop {} {\bibfield
		{journal} {\bibinfo  {journal} {Int. J. Heat Fluid Fl.}\ }\textbf {\bibinfo
			{volume} {21}},\ \bibinfo {pages} {359--364} (\bibinfo {year}
		{2000})}\BibitemShut {NoStop}%
	\bibitem [{\citenamefont {Sieber}\ \emph {et~al.}(2016)\citenamefont {Sieber},
		\citenamefont {Paschereit},\ and\ \citenamefont {Oberleithner}}]{sieberSPOD}%
	\BibitemOpen
	\bibfield  {author} {\bibinfo {author} {\bibfnamefont {M.}~\bibnamefont
			{Sieber}}, \bibinfo {author} {\bibfnamefont {C.~O.}\ \bibnamefont
			{Paschereit}}, \ and\ \bibinfo {author} {\bibfnamefont {K.}~\bibnamefont
			{Oberleithner}},\ }\bibfield  {title} {\enquote {\bibinfo {title} {Spectral
				proper orthogonal decomposition},}\ }\href@noop {} {\bibfield  {journal}
		{\bibinfo  {journal} {J. Fluid Mech.}\ }\textbf {\bibinfo {volume} {792}},\
		\bibinfo {pages} {798--828} (\bibinfo {year} {2016})}\BibitemShut {NoStop}%
	\bibitem [{\citenamefont {Cavalieri}\ \emph {et~al.}(2013)\citenamefont
		{Cavalieri}, \citenamefont {Rodriguez}, \citenamefont {Jordan}, \citenamefont
		{Colonius},\ and\ \citenamefont {Gervais}}]{cavalieri2013wavepackets}%
	\BibitemOpen
	\bibfield  {author} {\bibinfo {author} {\bibfnamefont {A.~V.~G.}\
			\bibnamefont {Cavalieri}}, \bibinfo {author} {\bibfnamefont {D.}~\bibnamefont
			{Rodriguez}}, \bibinfo {author} {\bibfnamefont {P.}~\bibnamefont {Jordan}},
		\bibinfo {author} {\bibfnamefont {T.}~\bibnamefont {Colonius}}, \ and\
		\bibinfo {author} {\bibfnamefont {Y.}~\bibnamefont {Gervais}},\ }\bibfield
	{title} {\enquote {\bibinfo {title} {Wavepackets in the velocity field of
				turbulent jets},}\ }\href@noop {} {\bibfield  {journal} {\bibinfo  {journal}
			{‎J. Fluid Mech.}\ }\textbf {\bibinfo {volume} {730}},\ \bibinfo {pages}
		{559--592} (\bibinfo {year} {2013})}\BibitemShut {NoStop}%
	\bibitem [{\citenamefont {Rodriguez}\ \emph {et~al.}(2015)\citenamefont
		{Rodriguez}, \citenamefont {Cavalieri}, \citenamefont {Colonius},\ and\
		\citenamefont {Jordan}}]{rodriguez2015}%
	\BibitemOpen
	\bibfield  {author} {\bibinfo {author} {\bibfnamefont {D.}~\bibnamefont
			{Rodriguez}}, \bibinfo {author} {\bibfnamefont {A.~V.~G.}\ \bibnamefont
			{Cavalieri}}, \bibinfo {author} {\bibfnamefont {T.}~\bibnamefont {Colonius}},
		\ and\ \bibinfo {author} {\bibfnamefont {P.}~\bibnamefont {Jordan}},\
	}\bibfield  {title} {\enquote {\bibinfo {title} {A study of linear wavepacket
				models for subsonic turbulent jets using local eigenmode decomposition of
				\mbox{PIV} data},}\ }\href@noop {} {\bibfield  {journal} {\bibinfo  {journal}
			{Eur. J. Mech. B/Fluids}\ }\textbf {\bibinfo {volume} {49B}},\ \bibinfo
		{pages} {308--321} (\bibinfo {year} {2015})}\BibitemShut {NoStop}%
	\bibitem [{\citenamefont {Garnaud}\ \emph
		{et~al.}(2013{\natexlab{a}})\citenamefont {Garnaud}, \citenamefont
		{Lesshafft}, \citenamefont {Schmid},\ and\ \citenamefont
		{Huerre}}]{garnaud2013preferred}%
	\BibitemOpen
	\bibfield  {author} {\bibinfo {author} {\bibfnamefont {X.}~\bibnamefont
			{Garnaud}}, \bibinfo {author} {\bibfnamefont {L.}~\bibnamefont {Lesshafft}},
		\bibinfo {author} {\bibfnamefont {P.~J.}\ \bibnamefont {Schmid}}, \ and\
		\bibinfo {author} {\bibfnamefont {P.}~\bibnamefont {Huerre}},\ }\bibfield
	{title} {\enquote {\bibinfo {title} {The preferred mode of incompressible
				jets: linear frequency response analysis},}\ }\href@noop {} {\bibfield
		{journal} {\bibinfo  {journal} {‎J. Fluid Mech.}\ }\textbf {\bibinfo
			{volume} {716}},\ \bibinfo {pages} {189--202} (\bibinfo {year}
		{2013}{\natexlab{a}})}\BibitemShut {NoStop}%
	\bibitem [{\citenamefont {Garnaud}\ \emph
		{et~al.}(2013{\natexlab{b}})\citenamefont {Garnaud}, \citenamefont
		{Sandberg},\ and\ \citenamefont {Lesshafft}}]{garnaud2013global}%
	\BibitemOpen
	\bibfield  {author} {\bibinfo {author} {\bibfnamefont {X.}~\bibnamefont
			{Garnaud}}, \bibinfo {author} {\bibfnamefont {R.~D.}\ \bibnamefont
			{Sandberg}}, \ and\ \bibinfo {author} {\bibfnamefont {L.}~\bibnamefont
			{Lesshafft}},\ }\bibfield  {title} {\enquote {\bibinfo {title} {Global
				response to forcing in a subsonic jet: instability wavepackets and acoustic
				radiation},}\ }\href@noop {} {\bibfield  {journal} {\bibinfo  {journal} {AIAA
				Paper 2013--4633}\ } (\bibinfo {year} {2013}{\natexlab{b}})}\BibitemShut
	{NoStop}%
	\bibitem [{\citenamefont {Jeun}\ \emph {et~al.}(2016)\citenamefont {Jeun},
		\citenamefont {Nichols},\ and\ \citenamefont {Jovanovi\'c}}]{jeun2016input}%
	\BibitemOpen
	\bibfield  {author} {\bibinfo {author} {\bibfnamefont {J.}~\bibnamefont
			{Jeun}}, \bibinfo {author} {\bibfnamefont {J.~W.}\ \bibnamefont {Nichols}}, \
		and\ \bibinfo {author} {\bibfnamefont {M.~R.}\ \bibnamefont {Jovanovi\'c}},\
	}\bibfield  {title} {\enquote {\bibinfo {title} {Input-output analysis of
				high-speed axisymmetric isothermal jet noise},}\ }\href@noop {} {\bibfield
		{journal} {\bibinfo  {journal} {Phys. Fluids}\ }\textbf {\bibinfo {volume}
			{28}},\ \bibinfo {pages} {047101} (\bibinfo {year} {2016})}\BibitemShut
	{NoStop}%
	\bibitem [{\citenamefont {Semeraro}\ \emph
		{et~al.}(2016{\natexlab{a}})\citenamefont {Semeraro}, \citenamefont
		{Lesshafft}, \citenamefont {Jaunet},\ and\ \citenamefont
		{Jordan}}]{semeraro2016modeling}%
	\BibitemOpen
	\bibfield  {author} {\bibinfo {author} {\bibfnamefont {O.}~\bibnamefont
			{Semeraro}}, \bibinfo {author} {\bibfnamefont {L.}~\bibnamefont {Lesshafft}},
		\bibinfo {author} {\bibfnamefont {V.}~\bibnamefont {Jaunet}}, \ and\ \bibinfo
		{author} {\bibfnamefont {P.}~\bibnamefont {Jordan}},\ }\bibfield  {title}
	{\enquote {\bibinfo {title} {Modeling of coherent structures in a turbulent
				jet as global linear instability wavepackets: Theory and experiment},}\
	}\href@noop {} {\bibfield  {journal} {\bibinfo  {journal} {Int. J. Heat Fluid
				Fl.}\ }\textbf {\bibinfo {volume} {62}},\ \bibinfo {pages} {24--32} (\bibinfo
		{year} {2016}{\natexlab{a}})}\BibitemShut {NoStop}%
	\bibitem [{\citenamefont {Semeraro}\ \emph
		{et~al.}(2016{\natexlab{b}})\citenamefont {Semeraro}, \citenamefont {Jaunet},
		\citenamefont {Jordan}, \citenamefont {Cavalieri},\ and\ \citenamefont
		{Lesshafft}}]{semeraro2016stochastic}%
	\BibitemOpen
	\bibfield  {author} {\bibinfo {author} {\bibfnamefont {O.}~\bibnamefont
			{Semeraro}}, \bibinfo {author} {\bibfnamefont {V.}~\bibnamefont {Jaunet}},
		\bibinfo {author} {\bibfnamefont {P.}~\bibnamefont {Jordan}}, \bibinfo
		{author} {\bibfnamefont {A.~V.~G.}\ \bibnamefont {Cavalieri}}, \ and\
		\bibinfo {author} {\bibfnamefont {L.}~\bibnamefont {Lesshafft}},\ }\bibfield
	{title} {\enquote {\bibinfo {title} {Stochastic and harmonic optimal forcing
				in subsonic jets},}\ }\href@noop {} {\bibfield  {journal} {\bibinfo
			{journal} {AIAA Paper 2016-2935}\ } (\bibinfo {year}
		{2016}{\natexlab{b}})}\BibitemShut {NoStop}%
	\bibitem [{\citenamefont {Towne}\ \emph {et~al.}(2018)\citenamefont {Towne},
		\citenamefont {Schmidt},\ and\ \citenamefont {Colonius}}]{towne2018}%
	\BibitemOpen
	\bibfield  {author} {\bibinfo {author} {\bibfnamefont {A.}~\bibnamefont
			{Towne}}, \bibinfo {author} {\bibfnamefont {O.~T.}\ \bibnamefont {Schmidt}},
		\ and\ \bibinfo {author} {\bibfnamefont {T.}~\bibnamefont {Colonius}},\
	}\bibfield  {title} {\enquote {\bibinfo {title} {Spectral proper orthogonal
				decomposition and its relationship to dynamic mode decomposition and
				resolvent analysis},}\ }\href@noop {} {\bibfield  {journal} {\bibinfo
			{journal} {J. Fluid Mech.}\ }\textbf {\bibinfo {volume} {847}},\ \bibinfo
		{pages} {821--867} (\bibinfo {year} {2018})}\BibitemShut {NoStop}%
	\bibitem [{\citenamefont {Schmidt}\ \emph {et~al.}(2018)\citenamefont
		{Schmidt}, \citenamefont {Towne}, \citenamefont {Rigas}, \citenamefont
		{Colonius},\ and\ \citenamefont {Br\`es}}]{schmidt2018}%
	\BibitemOpen
	\bibfield  {author} {\bibinfo {author} {\bibfnamefont {O.~T.}\ \bibnamefont
			{Schmidt}}, \bibinfo {author} {\bibfnamefont {A.}~\bibnamefont {Towne}},
		\bibinfo {author} {\bibfnamefont {G.}~\bibnamefont {Rigas}}, \bibinfo
		{author} {\bibfnamefont {T.}~\bibnamefont {Colonius}}, \ and\ \bibinfo
		{author} {\bibfnamefont {G.~A.}\ \bibnamefont {Br\`es}},\ }\bibfield  {title}
	{\enquote {\bibinfo {title} {Spectral analysis of jet turbulence},}\
	}\href@noop {} {\bibfield  {journal} {\bibinfo  {journal} {J. Fluid Mech.}\
		}\textbf {\bibinfo {volume} {855}},\ \bibinfo {pages} {953--982} (\bibinfo
		{year} {2018})}\BibitemShut {NoStop}%
	\bibitem [{\citenamefont {Farrell}\ and\ \citenamefont
		{Ioannou}(1996)}]{farrell1996generalized}%
	\BibitemOpen
	\bibfield  {author} {\bibinfo {author} {\bibfnamefont {B.~F.}\ \bibnamefont
			{Farrell}}\ and\ \bibinfo {author} {\bibfnamefont {P.~J.}\ \bibnamefont
			{Ioannou}},\ }\bibfield  {title} {\enquote {\bibinfo {title} {Generalized
				stability theory. {P}art {I}: Autonomous operators},}\ }\href@noop {}
	{\bibfield  {journal} {\bibinfo  {journal} {J. Atmospheric Sci.}\ }\textbf
		{\bibinfo {volume} {53}},\ \bibinfo {pages} {2025--2040} (\bibinfo {year}
		{1996})}\BibitemShut {NoStop}%
	\bibitem [{\citenamefont {Alizard}\ \emph {et~al.}(2009)\citenamefont
		{Alizard}, \citenamefont {Cherubini},\ and\ \citenamefont
		{Robinet}}]{alizard2009sensitivity}%
	\BibitemOpen
	\bibfield  {author} {\bibinfo {author} {\bibfnamefont {F.}~\bibnamefont
			{Alizard}}, \bibinfo {author} {\bibfnamefont {S.}~\bibnamefont {Cherubini}},
		\ and\ \bibinfo {author} {\bibfnamefont {J.-C.}\ \bibnamefont {Robinet}},\
	}\bibfield  {title} {\enquote {\bibinfo {title} {Sensitivity and optimal
				forcing response in separated boundary layer flows},}\ }\href@noop {}
	{\bibfield  {journal} {\bibinfo  {journal} {Phys. Fluids}\ }\textbf {\bibinfo
			{volume} {21}},\ \bibinfo {pages} {064108} (\bibinfo {year}
		{2009})}\BibitemShut {NoStop}%
	\bibitem [{\citenamefont {Monokrousos}\ \emph {et~al.}(2010)\citenamefont
		{Monokrousos}, \citenamefont {{\AA}kervik}, \citenamefont {Brandt},\ and\
		\citenamefont {Henningson}}]{monokrousos2010global}%
	\BibitemOpen
	\bibfield  {author} {\bibinfo {author} {\bibfnamefont {A.}~\bibnamefont
			{Monokrousos}}, \bibinfo {author} {\bibfnamefont {E.}~\bibnamefont
			{{\AA}kervik}}, \bibinfo {author} {\bibfnamefont {L.}~\bibnamefont {Brandt}},
		\ and\ \bibinfo {author} {\bibfnamefont {D.~S.}\ \bibnamefont {Henningson}},\
	}\bibfield  {title} {\enquote {\bibinfo {title} {Global three-dimensional
				optimal disturbances in the {B}lasius boundary-layer flow using
				time-steppers},}\ }\href@noop {} {\bibfield  {journal} {\bibinfo  {journal}
			{‎J. Fluid Mech.}\ }\textbf {\bibinfo {volume} {650}},\ \bibinfo {pages}
		{181--214} (\bibinfo {year} {2010})}\BibitemShut {NoStop}%
	\bibitem [{\citenamefont {Sipp}\ and\ \citenamefont
		{Marquet}(2013)}]{sipp2013characterization}%
	\BibitemOpen
	\bibfield  {author} {\bibinfo {author} {\bibfnamefont {D.}~\bibnamefont
			{Sipp}}\ and\ \bibinfo {author} {\bibfnamefont {O.}~\bibnamefont {Marquet}},\
	}\bibfield  {title} {\enquote {\bibinfo {title} {Characterization of noise
				amplifiers with global singular modes: the case of the leading-edge
				flat-plate boundary layer},}\ }\href@noop {} {\bibfield  {journal} {\bibinfo
			{journal} {Theor. Comp. Fluid Dyn.}\ }\textbf {\bibinfo {volume} {27}},\
		\bibinfo {pages} {617--635} (\bibinfo {year} {2013})}\BibitemShut {NoStop}%
	\bibitem [{\citenamefont {Dergham}\ \emph {et~al.}(2013)\citenamefont
		{Dergham}, \citenamefont {Sipp},\ and\ \citenamefont
		{Robinet}}]{dergham2013stochastic}%
	\BibitemOpen
	\bibfield  {author} {\bibinfo {author} {\bibfnamefont {G.}~\bibnamefont
			{Dergham}}, \bibinfo {author} {\bibfnamefont {D.}~\bibnamefont {Sipp}}, \
		and\ \bibinfo {author} {\bibfnamefont {J.-C.}\ \bibnamefont {Robinet}},\
	}\bibfield  {title} {\enquote {\bibinfo {title} {Stochastic dynamics and
				model reduction of amplifier flows: the backward facing step flow},}\
	}\href@noop {} {\bibfield  {journal} {\bibinfo  {journal} {‎J. Fluid
				Mech.}\ }\textbf {\bibinfo {volume} {719}},\ \bibinfo {pages} {406--430}
		(\bibinfo {year} {2013})}\BibitemShut {NoStop}%
	\bibitem [{\citenamefont {Boujo}\ and\ \citenamefont
		{Gallaire}(2015)}]{boujo2015sensitivity}%
	\BibitemOpen
	\bibfield  {author} {\bibinfo {author} {\bibfnamefont {E.}~\bibnamefont
			{Boujo}}\ and\ \bibinfo {author} {\bibfnamefont {F.}~\bibnamefont
			{Gallaire}},\ }\bibfield  {title} {\enquote {\bibinfo {title} {Sensitivity
				and open-loop control of stochastic response in a noise amplifier flow: the
				backward-facing step},}\ }\href@noop {} {\bibfield  {journal} {\bibinfo
			{journal} {‎J. Fluid Mech.}\ }\textbf {\bibinfo {volume} {762}},\ \bibinfo
		{pages} {361--392} (\bibinfo {year} {2015})}\BibitemShut {NoStop}%
	\bibitem [{\citenamefont {Beneddine}\ \emph {et~al.}(2016)\citenamefont
		{Beneddine}, \citenamefont {Sipp}, \citenamefont {Arnault}, \citenamefont
		{Dandois},\ and\ \citenamefont {Lesshafft}}]{beneddine2016conditions}%
	\BibitemOpen
	\bibfield  {author} {\bibinfo {author} {\bibfnamefont {S.}~\bibnamefont
			{Beneddine}}, \bibinfo {author} {\bibfnamefont {D.}~\bibnamefont {Sipp}},
		\bibinfo {author} {\bibfnamefont {A.}~\bibnamefont {Arnault}}, \bibinfo
		{author} {\bibfnamefont {J.}~\bibnamefont {Dandois}}, \ and\ \bibinfo
		{author} {\bibfnamefont {L.}~\bibnamefont {Lesshafft}},\ }\bibfield  {title}
	{\enquote {\bibinfo {title} {Conditions for validity of mean flow stability
				analysis},}\ }\href@noop {} {\bibfield  {journal} {\bibinfo  {journal} {‎J.
				Fluid Mech.}\ }\textbf {\bibinfo {volume} {798}},\ \bibinfo {pages}
		{485--504} (\bibinfo {year} {2016})}\BibitemShut {NoStop}%
	\bibitem [{\citenamefont {Towne}\ \emph {et~al.}(2015)\citenamefont {Towne},
		\citenamefont {Colonius}, \citenamefont {Jordan}, \citenamefont {Cavalieri},\
		and\ \citenamefont {Br{\`e}s}}]{towne2015stochastic}%
	\BibitemOpen
	\bibfield  {author} {\bibinfo {author} {\bibfnamefont {A.}~\bibnamefont
			{Towne}}, \bibinfo {author} {\bibfnamefont {T.}~\bibnamefont {Colonius}},
		\bibinfo {author} {\bibfnamefont {P.}~\bibnamefont {Jordan}}, \bibinfo
		{author} {\bibfnamefont {A.~V.~G.}\ \bibnamefont {Cavalieri}}, \ and\
		\bibinfo {author} {\bibfnamefont {G.~A.}\ \bibnamefont {Br{\`e}s}},\
	}\bibfield  {title} {\enquote {\bibinfo {title} {Stochastic and nonlinear
				forcing of wavepackets in a {M}ach 0.9 jet},}\ }\href@noop {} {\bibfield
		{journal} {\bibinfo  {journal} {AIAA Paper 2015-2217}\ } (\bibinfo {year}
		{2015})}\BibitemShut {NoStop}%
	\bibitem [{\citenamefont {Cavalieri}\ \emph {et~al.}(2019)\citenamefont
		{Cavalieri}, \citenamefont {Jordan},\ and\ \citenamefont
		{Lesshafft}}]{cavalieri2019amr}%
	\BibitemOpen
	\bibfield  {author} {\bibinfo {author} {\bibfnamefont {A.~V.~G.}\
			\bibnamefont {Cavalieri}}, \bibinfo {author} {\bibfnamefont {P.}~\bibnamefont
			{Jordan}}, \ and\ \bibinfo {author} {\bibfnamefont {L.}~\bibnamefont
			{Lesshafft}},\ }\bibfield  {title} {\enquote {\bibinfo {title} {Wave-packet
				models for jet dynamics and sound radiation},}\ }\href {\doibase
		doi:10.1115/1.4042736} {\bibfield  {journal} {\bibinfo  {journal} {Appl.
				Mech. Rev.}\ } (\bibinfo {year} {2019}),\ doi:10.1115/1.4042736}\BibitemShut
	{NoStop}%
	\bibitem [{\citenamefont {Jaunet}\ \emph {et~al.}(2017)\citenamefont {Jaunet},
		\citenamefont {Jordan},\ and\ \citenamefont {Cavalieri}}]{jaunet2017prf}%
	\BibitemOpen
	\bibfield  {author} {\bibinfo {author} {\bibfnamefont {V.}~\bibnamefont
			{Jaunet}}, \bibinfo {author} {\bibfnamefont {P.}~\bibnamefont {Jordan}}, \
		and\ \bibinfo {author} {\bibfnamefont {A.~V.~G.}\ \bibnamefont {Cavalieri}},\
	}\bibfield  {title} {\enquote {\bibinfo {title} {Two-point coherence of wave
				packets in turbulent jets},}\ }\href {\doibase
		10.1103/PhysRevFluids.2.024604} {\bibfield  {journal} {\bibinfo  {journal}
			{Phys. Rev. Fluids}\ }\textbf {\bibinfo {volume} {2}},\ \bibinfo {pages}
		{024604} (\bibinfo {year} {2017})}\BibitemShut {NoStop}%
	\bibitem [{\citenamefont {Br{\`e}s}\ \emph {et~al.}(2018)\citenamefont
		{Br{\`e}s}, \citenamefont {Jordan}, \citenamefont {Jaunet}, \citenamefont
		{{Le Rallic}}, \citenamefont {Cavalieri}, \citenamefont {Towne},
		\citenamefont {Lele}, \citenamefont {Colonius},\ and\ \citenamefont
		{Schmidt}}]{bres2018importance}%
	\BibitemOpen
	\bibfield  {author} {\bibinfo {author} {\bibfnamefont {G.~A.}\ \bibnamefont
			{Br{\`e}s}}, \bibinfo {author} {\bibfnamefont {P.}~\bibnamefont {Jordan}},
		\bibinfo {author} {\bibfnamefont {V.}~\bibnamefont {Jaunet}}, \bibinfo
		{author} {\bibfnamefont {M.}~\bibnamefont {{Le Rallic}}}, \bibinfo {author}
		{\bibfnamefont {A.~V.~G.}\ \bibnamefont {Cavalieri}}, \bibinfo {author}
		{\bibfnamefont {A.}~\bibnamefont {Towne}}, \bibinfo {author} {\bibfnamefont
			{S.~K.}\ \bibnamefont {Lele}}, \bibinfo {author} {\bibfnamefont
			{T.}~\bibnamefont {Colonius}}, \ and\ \bibinfo {author} {\bibfnamefont
			{O.~T.}\ \bibnamefont {Schmidt}},\ }\bibfield  {title} {\enquote {\bibinfo
			{title} {Importance of the nozzle-exit boundary-layer state in subsonic
				turbulent jets},}\ }\href@noop {} {\bibfield  {journal} {\bibinfo  {journal}
			{J. Fluid Mech.}\ }\textbf {\bibinfo {volume} {851}},\ \bibinfo {pages}
		{83--124} (\bibinfo {year} {2018})}\BibitemShut {NoStop}%
	\bibitem {meanflow}%
	See Supplemental Material at [URL will be inserted by publisher] for the mean flow file.
	\bibitem [{\citenamefont {Pope}(2000)}]{pope2000turbulent}%
	\BibitemOpen
	\bibfield  {author} {\bibinfo {author} {\bibfnamefont {S.~B.}\ \bibnamefont
			{Pope}},\ }\href@noop {} {\emph {\bibinfo {title} {Turbulent flows}}}\
	(\bibinfo  {publisher} {Cambridge {U}niversity {P}ress},\ \bibinfo {year}
	{2000})\BibitemShut {NoStop}%
	\bibitem [{\citenamefont {Garnaud}(2012)}]{garnaud2012modes}%
	\BibitemOpen
	\bibfield  {author} {\bibinfo {author} {\bibfnamefont {X.}~\bibnamefont
			{Garnaud}},\ }\emph {\bibinfo {title} {Modes, transient dynamics and forced
			response of circular jets}},\ \href@noop {} {Ph.D. thesis},\ \bibinfo
	{school} {Ecole polytechnique, Palaiseau, France} (\bibinfo {year} {2012})\BibitemShut {NoStop}%
	\bibitem [{\citenamefont {Sandberg}(2007)}]{sandberg2007governing}%
	\BibitemOpen
	\bibfield  {author} {\bibinfo {author} {\bibfnamefont {R.~D.}\ \bibnamefont
			{Sandberg}},\ }\bibfield  {title} {\enquote {\bibinfo {title} {Governing
				equations for a new compressible {N}avier-{S}tokes solver in general
				cylindrical coordinates},}\ }\href@noop {} {\bibfield  {journal} {\bibinfo
			{journal} {Monograph No. AFM-07/07, School of Engineering Sciences,
				University of Southampton}\ } (\bibinfo {year} {2007})}\BibitemShut {NoStop}%
	\bibitem [{\citenamefont {Reynolds}\ and\ \citenamefont
		{Hussain}(1972)}]{reynolds1972mechanics}%
	\BibitemOpen
	\bibfield  {author} {\bibinfo {author} {\bibfnamefont {W.~C.}\ \bibnamefont
			{Reynolds}}\ and\ \bibinfo {author} {\bibfnamefont {A.~K.~M.~F.}\
			\bibnamefont {Hussain}},\ }\bibfield  {title} {\enquote {\bibinfo {title}
			{The mechanics of an organized wave in turbulent shear flow. {P}art 3.
				{T}heoretical models and comparisons with experiments},}\ }\href@noop {}
	{\bibfield  {journal} {\bibinfo  {journal} {‎J. Fluid Mech.}\ }\textbf
		{\bibinfo {volume} {54}},\ \bibinfo {pages} {263--288} (\bibinfo {year}
		{1972})}\BibitemShut {NoStop}%
	\bibitem [{\citenamefont {McKeon}\ and\ \citenamefont
		{Sharma}(2010)}]{mckeon2010critical}%
	\BibitemOpen
	\bibfield  {author} {\bibinfo {author} {\bibfnamefont {B.~J.}\ \bibnamefont
			{McKeon}}\ and\ \bibinfo {author} {\bibfnamefont {A.~S.}\ \bibnamefont
			{Sharma}},\ }\bibfield  {title} {\enquote {\bibinfo {title} {A critical-layer
				framework for turbulent pipe flow},}\ }\href@noop {} {\bibfield  {journal}
		{\bibinfo  {journal} {‎J. Fluid Mech.}\ }\textbf {\bibinfo {volume}
			{658}},\ \bibinfo {pages} {336--382} (\bibinfo {year} {2010})}\BibitemShut
	{NoStop}%
	\bibitem [{\citenamefont {Tammisola}\ and\ \citenamefont
		{Juniper}(2016)}]{tammisola2016coherent}%
	\BibitemOpen
	\bibfield  {author} {\bibinfo {author} {\bibfnamefont {O.}~\bibnamefont
			{Tammisola}}\ and\ \bibinfo {author} {\bibfnamefont {M.~P.}\ \bibnamefont
			{Juniper}},\ }\bibfield  {title} {\enquote {\bibinfo {title} {Coherent
				structures in a swirl injector at {R}e=4800 by nonlinear simulations and
				linear global modes},}\ }\href@noop {} {\bibfield  {journal} {\bibinfo
			{journal} {‎J. Fluid Mech.}\ }\textbf {\bibinfo {volume} {792}},\ \bibinfo
		{pages} {620--657} (\bibinfo {year} {2016})}\BibitemShut {NoStop}%
	\bibitem [{\citenamefont {Oberleithner}\ \emph {et~al.}(2015)\citenamefont
		{Oberleithner}, \citenamefont {St{\"o}hr}, \citenamefont {Im}, \citenamefont
		{Arndt},\ and\ \citenamefont {Steinberg}}]{oberleithner2015formation}%
	\BibitemOpen
	\bibfield  {author} {\bibinfo {author} {\bibfnamefont {K.}~\bibnamefont
			{Oberleithner}}, \bibinfo {author} {\bibfnamefont {M.}~\bibnamefont
			{St{\"o}hr}}, \bibinfo {author} {\bibfnamefont {S.~H.}\ \bibnamefont {Im}},
		\bibinfo {author} {\bibfnamefont {C.~M.}\ \bibnamefont {Arndt}}, \ and\
		\bibinfo {author} {\bibfnamefont {A.~M.}\ \bibnamefont {Steinberg}},\
	}\bibfield  {title} {\enquote {\bibinfo {title} {Formation and flame-induced
				suppression of the precessing vortex core in a swirl combustor: experiments
				and linear stability analysis},}\ }\href@noop {} {\bibfield  {journal}
		{\bibinfo  {journal} {Combust. Flame}\ }\textbf {\bibinfo {volume} {162}},\
		\bibinfo {pages} {3100--3114} (\bibinfo {year} {2015})}\BibitemShut {NoStop}%
	\bibitem [{\citenamefont {Semeraro}\ \emph
		{et~al.}(2016{\natexlab{c}})\citenamefont {Semeraro}, \citenamefont
		{Lesshafft},\ and\ \citenamefont {Sandberg}}]{semeraroSandbergStockholm}%
	\BibitemOpen
	\bibfield  {author} {\bibinfo {author} {\bibfnamefont {O.}~\bibnamefont
			{Semeraro}}, \bibinfo {author} {\bibfnamefont {L.}~\bibnamefont {Lesshafft}},
		\ and\ \bibinfo {author} {\bibfnamefont {R.~D.}\ \bibnamefont {Sandberg}},\
	}\bibfield  {title} {\enquote {\bibinfo {title} {Can jet noise be predicted
				using linear instability wavepackets?}}\ }in\ \href@noop {} {\emph {\bibinfo
			{booktitle} {Proceedings of the 5th International Conference on Jets, Wakes
				and Separated Flows}}}\ (\bibinfo {organization} {Springer},\ \bibinfo {year}
	{2016})\ pp.\ \bibinfo {pages} {413--418}\BibitemShut {NoStop}%
	\bibitem [{\citenamefont {{\AA}kervik}\ \emph {et~al.}(2008)\citenamefont
		{{\AA}kervik}, \citenamefont {Ehrenstein}, \citenamefont {Gallaire},\ and\
		\citenamefont {Henningson}}]{aakervik2008global}%
	\BibitemOpen
	\bibfield  {author} {\bibinfo {author} {\bibfnamefont {E.}~\bibnamefont
			{{\AA}kervik}}, \bibinfo {author} {\bibfnamefont {U.}~\bibnamefont
			{Ehrenstein}}, \bibinfo {author} {\bibfnamefont {F.}~\bibnamefont
			{Gallaire}}, \ and\ \bibinfo {author} {\bibfnamefont {D.~S.}\ \bibnamefont
			{Henningson}},\ }\bibfield  {title} {\enquote {\bibinfo {title} {Global
				two-dimensional stability measures of the flat plate boundary-layer flow},}\
	}\href@noop {} {\bibfield  {journal} {\bibinfo  {journal} {Eur. J. Mech.
				B/Fluids}\ }\textbf {\bibinfo {volume} {27}},\ \bibinfo {pages} {501--513}
		(\bibinfo {year} {2008})}\BibitemShut {NoStop}%
	\bibitem [{\citenamefont {Huerre}\ and\ \citenamefont
		{Monkewitz}(1990)}]{huerre1990local}%
	\BibitemOpen
	\bibfield  {author} {\bibinfo {author} {\bibfnamefont {P.}~\bibnamefont
			{Huerre}}\ and\ \bibinfo {author} {\bibfnamefont {P.}~\bibnamefont
			{Monkewitz}},\ }\bibfield  {title} {\enquote {\bibinfo {title} {Local and
				global instabilities in spatially developing flows},}\ }\href@noop {}
	{\bibfield  {journal} {\bibinfo  {journal} {Annu. Rev. Fluid Mech.}\ }\textbf
		{\bibinfo {volume} {22}},\ \bibinfo {pages} {473--537} (\bibinfo {year}
		{1990})}\BibitemShut {NoStop}%
	\bibitem [{\citenamefont {Garnaud}\ \emph
		{et~al.}(2013{\natexlab{c}})\citenamefont {Garnaud}, \citenamefont
		{Lesshafft}, \citenamefont {Schmid},\ and\ \citenamefont
		{Huerre}}]{garnaud2013modal}%
	\BibitemOpen
	\bibfield  {author} {\bibinfo {author} {\bibfnamefont {X.}~\bibnamefont
			{Garnaud}}, \bibinfo {author} {\bibfnamefont {L.}~\bibnamefont {Lesshafft}},
		\bibinfo {author} {\bibfnamefont {P.~J.}\ \bibnamefont {Schmid}}, \ and\
		\bibinfo {author} {\bibfnamefont {P.}~\bibnamefont {Huerre}},\ }\bibfield
	{title} {\enquote {\bibinfo {title} {Modal and transient dynamics of jet
				flows},}\ }\href@noop {} {\bibfield  {journal} {\bibinfo  {journal} {Phys.
				Fluids}\ }\textbf {\bibinfo {volume} {25}},\ \bibinfo {pages} {044103}
		(\bibinfo {year} {2013}{\natexlab{c}})}\BibitemShut {NoStop}%
	\bibitem [{\citenamefont {Chu}(1965)}]{chu1965}%
	\BibitemOpen
	\bibfield  {author} {\bibinfo {author} {\bibfnamefont {B.-T.}\ \bibnamefont
			{Chu}},\ }\bibfield  {title} {\enquote {\bibinfo {title} {On the energy
				transfer to small disturbances in fluid flow ({P}art {I})},}\ }\href@noop {}
	{\bibfield  {journal} {\bibinfo  {journal} {Acta Mech.}\ }\textbf {\bibinfo
			{volume} {1}},\ \bibinfo {pages} {215--234} (\bibinfo {year}
		{1965})}\BibitemShut {NoStop}%
	\bibitem [{\citenamefont {Hernandez}\ \emph {et~al.}(2005)\citenamefont
		{Hernandez}, \citenamefont {Roman},\ and\ \citenamefont
		{Vidal}}]{hernandez2005slepc}%
	\BibitemOpen
	\bibfield  {author} {\bibinfo {author} {\bibfnamefont {V.}~\bibnamefont
			{Hernandez}}, \bibinfo {author} {\bibfnamefont {J.~E.}\ \bibnamefont
			{Roman}}, \ and\ \bibinfo {author} {\bibfnamefont {V.}~\bibnamefont
			{Vidal}},\ }\bibfield  {title} {\enquote {\bibinfo {title} {{SLEPc}: A
				scalable and flexible toolkit for the solution of eigenvalue problems},}\
	}\href@noop {} {\bibfield  {journal} {\bibinfo  {journal} {ACM Transactions
				on Mathematical Software (TOMS)}\ }\textbf {\bibinfo {volume} {31}},\
		\bibinfo {pages} {351--362} (\bibinfo {year} {2005})}\BibitemShut {NoStop}%
	\bibitem [{\citenamefont {{Fosas de Pando}}\ \emph {et~al.}(2012)\citenamefont
		{{Fosas de Pando}}, \citenamefont {Sipp},\ and\ \citenamefont
		{Schmid}}]{de2012efficient}%
	\BibitemOpen
	\bibfield  {author} {\bibinfo {author} {\bibfnamefont {M.}~\bibnamefont
			{{Fosas de Pando}}}, \bibinfo {author} {\bibfnamefont {D.}~\bibnamefont
			{Sipp}}, \ and\ \bibinfo {author} {\bibfnamefont {P.~J.}\ \bibnamefont
			{Schmid}},\ }\bibfield  {title} {\enquote {\bibinfo {title} {Efficient
				evaluation of the direct and adjoint linearized dynamics from compressible
				flow solvers},}\ }\href@noop {} {\bibfield  {journal} {\bibinfo  {journal}
			{J. Comput. Phys.}\ }\textbf {\bibinfo {volume} {231}},\ \bibinfo {pages}
		{7739--7755} (\bibinfo {year} {2012})}\BibitemShut {NoStop}%
	\bibitem [{\citenamefont {Berland}\ \emph {et~al.}(2007)\citenamefont
		{Berland}, \citenamefont {Bogey}, \citenamefont {Marsden},\ and\
		\citenamefont {Bailly}}]{berland2007high}%
	\BibitemOpen
	\bibfield  {author} {\bibinfo {author} {\bibfnamefont {J.}~\bibnamefont
			{Berland}}, \bibinfo {author} {\bibfnamefont {C.}~\bibnamefont {Bogey}},
		\bibinfo {author} {\bibfnamefont {O.}~\bibnamefont {Marsden}}, \ and\
		\bibinfo {author} {\bibfnamefont {C.}~\bibnamefont {Bailly}},\ }\bibfield
	{title} {\enquote {\bibinfo {title} {High-order, low dispersive and low
				dissipative explicit schemes for multiple-scale and boundary problems},}\
	}\href@noop {} {\bibfield  {journal} {\bibinfo  {journal} {J. Comput. Phys.}\
		}\textbf {\bibinfo {volume} {224}},\ \bibinfo {pages} {637--662} (\bibinfo
		{year} {2007})}\BibitemShut {NoStop}%
	\bibitem [{\citenamefont {Poinsot}\ and\ \citenamefont
		{Lele}(1992)}]{poinsot1992}%
	\BibitemOpen
	\bibfield  {author} {\bibinfo {author} {\bibfnamefont {T.~J.}\ \bibnamefont
			{Poinsot}}\ and\ \bibinfo {author} {\bibfnamefont {S.~K.}\ \bibnamefont
			{Lele}},\ }\bibfield  {title} {\enquote {\bibinfo {title} {Boundary
				conditions for direct simulations of compressible viscous flows},}\
	}\href@noop {} {\bibfield  {journal} {\bibinfo  {journal} {J. Comput. Phys.}\
		}\textbf {\bibinfo {volume} {101}},\ \bibinfo {pages} {104--129} (\bibinfo
		{year} {1992})}\BibitemShut {NoStop}%
	\bibitem [{\citenamefont {Colonius}(2004)}]{coloniusARFM}%
	\BibitemOpen
	\bibfield  {author} {\bibinfo {author} {\bibfnamefont {T.}~\bibnamefont
			{Colonius}},\ }\bibfield  {title} {\enquote {\bibinfo {title} {Modeling
				artificial boundary conditions for compressible flow},}\ }\href@noop {}
	{\bibfield  {journal} {\bibinfo  {journal} {Annu. Rev. Fluid Mech.}\ }\textbf
		{\bibinfo {volume} {36}},\ \bibinfo {pages} {315--345} (\bibinfo {year}
		{2004})}\BibitemShut {NoStop}%
	\bibitem [{\citenamefont {Tissot}\ \emph
		{et~al.}(2017{\natexlab{a}})\citenamefont {Tissot}, \citenamefont {Zhang},
		\citenamefont {Lajus}, \citenamefont {Cavalieri}, \citenamefont {Jordan},\
		and\ \citenamefont {Colonius}}]{tissot2016sensitivity}%
	\BibitemOpen
	\bibfield  {author} {\bibinfo {author} {\bibfnamefont {G.}~\bibnamefont
			{Tissot}}, \bibinfo {author} {\bibfnamefont {M.}~\bibnamefont {Zhang}},
		\bibinfo {author} {\bibfnamefont {F.}~\bibnamefont {Lajus}}, \bibinfo
		{author} {\bibfnamefont {A.~V.~G.}\ \bibnamefont {Cavalieri}}, \bibinfo
		{author} {\bibfnamefont {P.}~\bibnamefont {Jordan}}, \ and\ \bibinfo {author}
		{\bibfnamefont {T.}~\bibnamefont {Colonius}},\ }\bibfield  {title} {\enquote
		{\bibinfo {title} {Sensitivity of wavepackets in jets to non-linear effects:
				the role of the critical layer},}\ }\href@noop {} {\bibfield  {journal}
		{\bibinfo  {journal} {‎J. Fluid Mech.}\ }\textbf {\bibinfo {volume}
			{811}},\ \bibinfo {pages} {95--137} (\bibinfo {year}
		{2017}{\natexlab{a}})}\BibitemShut {NoStop}%
	\bibitem [{\citenamefont {Tissot}\ \emph
		{et~al.}(2017{\natexlab{b}})\citenamefont {Tissot}, \citenamefont
		{Laj{\'u}s}, \citenamefont {Cavalieri},\ and\ \citenamefont
		{Jordan}}]{tissot2017wave}%
	\BibitemOpen
	\bibfield  {author} {\bibinfo {author} {\bibfnamefont {G.}~\bibnamefont
			{Tissot}}, \bibinfo {author} {\bibfnamefont {F.}~\bibnamefont {Laj{\'u}s}},
		\bibinfo {author} {\bibfnamefont {A.~V.~G.}\ \bibnamefont {Cavalieri}}, \
		and\ \bibinfo {author} {\bibfnamefont {P.}~\bibnamefont {Jordan}},\
	}\bibfield  {title} {\enquote {\bibinfo {title} {Wave packets and {O}rr
				mechanism in turbulent jets},}\ }\href@noop {} {\bibfield  {journal}
		{\bibinfo  {journal} {Phys. Rev. Fluids}\ }\textbf {\bibinfo {volume} {2}},\
		\bibinfo {pages} {093901} (\bibinfo {year} {2017}{\natexlab{b}})}\BibitemShut
	{NoStop}%
	\bibitem [{\citenamefont {Butler}\ and\ \citenamefont
		{Farrell}(1992)}]{butler1992three}%
	\BibitemOpen
	\bibfield  {author} {\bibinfo {author} {\bibfnamefont {K.~M.}\ \bibnamefont
			{Butler}}\ and\ \bibinfo {author} {\bibfnamefont {B.~F.}\ \bibnamefont
			{Farrell}},\ }\bibfield  {title} {\enquote {\bibinfo {title}
			{Three-dimensional optimal perturbations in viscous shear flow},}\
	}\href@noop {} {\bibfield  {journal} {\bibinfo  {journal} {Phys. Fluids A}\
		}\textbf {\bibinfo {volume} {4}},\ \bibinfo {pages} {1637--1650} (\bibinfo
		{year} {1992})}\BibitemShut {NoStop}%
	\bibitem [{\citenamefont {Jim\'enez}(2013)}]{jimenez2013}%
	\BibitemOpen
	\bibfield  {author} {\bibinfo {author} {\bibfnamefont {J.}~\bibnamefont
			{Jim\'enez}},\ }\bibfield  {title} {\enquote {\bibinfo {title} {How linear is
				wall-bounded turbulence?}}\ }\href@noop {} {\bibfield  {journal} {\bibinfo
			{journal} {Phys. Fluids}\ }\textbf {\bibinfo {volume} {25}},\ \bibinfo
		{pages} {110814} (\bibinfo {year} {2013})}\BibitemShut {NoStop}%
	\bibitem [{\citenamefont {Lesshafft}\ and\ \citenamefont
		{Huerre}(2007)}]{lesshafft2007linear}%
	\BibitemOpen
	\bibfield  {author} {\bibinfo {author} {\bibfnamefont {L.}~\bibnamefont
			{Lesshafft}}\ and\ \bibinfo {author} {\bibfnamefont {P.}~\bibnamefont
			{Huerre}},\ }\bibfield  {title} {\enquote {\bibinfo {title} {Linear impulse
				response in hot round jets},}\ }\href@noop {} {\bibfield  {journal} {\bibinfo
			{journal} {Phys. Fluids}\ }\textbf {\bibinfo {volume} {19}},\ \bibinfo
		{pages} {024102} (\bibinfo {year} {2007})}\BibitemShut {NoStop}%
	\bibitem [{\citenamefont {Le~Diz{\`e}s}\ \emph {et~al.}(1995)\citenamefont
		{Le~Diz{\`e}s}, \citenamefont {Monkewitz},\ and\ \citenamefont
		{Huerre}}]{le1995viscous}%
	\BibitemOpen
	\bibfield  {author} {\bibinfo {author} {\bibfnamefont {S.}~\bibnamefont
			{Le~Diz{\`e}s}}, \bibinfo {author} {\bibfnamefont {P.~A.}\ \bibnamefont
			{Monkewitz}}, \ and\ \bibinfo {author} {\bibfnamefont {P.}~\bibnamefont
			{Huerre}},\ }\bibfield  {title} {\enquote {\bibinfo {title} {Viscous
				structure of plane waves in spatially developing shear flows},}\ }\href@noop
	{} {\bibfield  {journal} {\bibinfo  {journal} {Phys.~Fluids}\ }\textbf
		{\bibinfo {volume} {7}},\ \bibinfo {pages} {1337--1347} (\bibinfo {year}
		{1995})}\BibitemShut {NoStop}%
	\bibitem [{\citenamefont {Lesshafft}(2018)}]{lesshafft2018artificial}%
	\BibitemOpen
	\bibfield  {author} {\bibinfo {author} {\bibfnamefont {L.}~\bibnamefont
			{Lesshafft}},\ }\bibfield  {title} {\enquote {\bibinfo {title} {Artificial
				eigenmodes in truncated flow domains},}\ }\href@noop {} {\bibfield  {journal}
		{\bibinfo  {journal} {Theor. Comput. Fluid Dyn.}\ }\textbf {\bibinfo {volume}
			{32}},\ \bibinfo {pages} {245--262} (\bibinfo {year} {2018})}\BibitemShut
	{NoStop}%
	\bibitem [{\citenamefont {Citriniti}\ and\ \citenamefont
		{George}(2000)}]{citriniti2000reconstruction}%
	\BibitemOpen
	\bibfield  {author} {\bibinfo {author} {\bibfnamefont {J.~H.}\ \bibnamefont
			{Citriniti}}\ and\ \bibinfo {author} {\bibfnamefont {W.~K.}\ \bibnamefont
			{George}},\ }\bibfield  {title} {\enquote {\bibinfo {title} {Reconstruction
				of the global velocity field in the axisymmetric mixing layer utilizing the
				proper orthogonal decomposition},}\ }\href@noop {} {\bibfield  {journal}
		{\bibinfo  {journal} {‎J. Fluid Mech.}\ }\textbf {\bibinfo {volume}
			{418}},\ \bibinfo {pages} {137--166} (\bibinfo {year} {2000})}\BibitemShut
	{NoStop}%
	\bibitem [{\citenamefont {Breakey}\ \emph {et~al.}(2013)\citenamefont
		{Breakey}, \citenamefont {Jordan}, \citenamefont {Cavalieri},\ and\
		\citenamefont {L{\'e}on}}]{breakey2013near}%
	\BibitemOpen
	\bibfield  {author} {\bibinfo {author} {\bibfnamefont {D.~E.}\ \bibnamefont
			{Breakey}}, \bibinfo {author} {\bibfnamefont {P.}~\bibnamefont {Jordan}},
		\bibinfo {author} {\bibfnamefont {A.~V.~G.}\ \bibnamefont {Cavalieri}}, \
		and\ \bibinfo {author} {\bibfnamefont {O.}~\bibnamefont {L{\'e}on}},\
	}\bibfield  {title} {\enquote {\bibinfo {title} {Near-field wavepackets and
				the far-field sound of a subsonic jet},}\ }\href@noop {} {\bibfield
		{journal} {\bibinfo  {journal} {AIAA Paper 2013-2083}\ } (\bibinfo {year}
		{2013})}\BibitemShut {NoStop}%
	\bibitem [{\citenamefont {Jordan}\ \emph {et~al.}(2017)\citenamefont {Jordan},
		\citenamefont {Zhang}, \citenamefont {Lehnasch},\ and\ \citenamefont
		{Cavalieri}}]{jordan2017modal}%
	\BibitemOpen
	\bibfield  {author} {\bibinfo {author} {\bibfnamefont {P.}~\bibnamefont
			{Jordan}}, \bibinfo {author} {\bibfnamefont {M.}~\bibnamefont {Zhang}},
		\bibinfo {author} {\bibfnamefont {G.}~\bibnamefont {Lehnasch}}, \ and\
		\bibinfo {author} {\bibfnamefont {A.~V.~G.}\ \bibnamefont {Cavalieri}},\
	}\bibfield  {title} {\enquote {\bibinfo {title} {Modal and non-modal linear
				wavepacket dynamics in turbulent jets},}\ }\href@noop {} {\bibfield
		{journal} {\bibinfo  {journal} {AIAA Paper 2017-3379}\ } (\bibinfo {year}
		{2017})}\BibitemShut {NoStop}%
\end{thebibliography}
\end{document}